\documentclass[10 pt, letterpaper]{article}

\usepackage{amsmath}
\usepackage{amssymb}
\usepackage{graphicx}

\newtheorem{proposition}{Proposition}
\newtheorem{thm}{Theorem}
\newtheorem{definition}{Definition}
\newtheorem{assumption}{Assumption}
\newtheorem{pf}{Proof}

\newtheorem{remark}{Remark}

\def\cF{\mathcal{F}}
\def\cB{\mathcal{B}}
\def\cP{\mathcal{P}}
\def\cX{\mathcal{X}}
\def\cK{\mathcal{K}}
\def\R{\mathbb{R}}
\def\X{\mathbb{X}}
\def\U{\mathbb{U}}

\begin{document}

\date{}
\title{On generalized terminal state constraints for\\ model predictive control
\thanks{This research has received funding from the European Union Seventh Framework
Programme (FP7/2007-2013) under grant agreement n. PIOF-GA-2009-252284 -
Marie Curie project ``Innovative Control, Identification and Estimation
Methodologies for Sustainable Energy Technologies'', and from grants AFOSR FA9550-09-1-0203 and NSF ECCS-0925637.}}

\author{Lorenzo Fagiano\thanks{Dip. di Automatica e Informatica, Politecnico di Torino,
Italy, and  Dept. of Mechanical Engineering, University of California at Santa Barbara, USA. E-mail: lorenzo.fagiano@polito.it.}$\,$ and Andrew R. Teel\thanks{Department of Electrical and Computer Engineering, University of California at Santa Barbara, Santa Barbara, CA, USA. E-mail: teel@ece.ucsb.edu.}
}
\maketitle

$\,$\\

\textbf{Abstract - This manuscript contains technical results related to a particular approach for the design of Model Predictive Control (MPC) laws. The approach, named ``generalized'' terminal state constraint,  induces the recursive feasibility of the underlying optimization problem and recursive satisfaction of state and input constraints, and it can be used for both tracking MPC (i.e. when the objective is to track a given steady state) and economic MPC (i.e. when the objective is to minimize a cost function which does not necessarily attains its minimum at a steady state). It is shown that the proposed technique provides, in general, a larger feasibility set with respect to existing approaches, given the same computational complexity. Moreover, a new receding horizon strategy is introduced, exploiting the generalized terminal state constraint. Under mild assumptions, the new strategy is guaranteed to converge in finite time, with arbitrarily good accuracy, to an MPC law with an optimally-chosen terminal state constraint, while still enjoying a larger feasibility set. The features of the new technique are illustrated by three examples.}

\section{Introduction}\label{S:Intro}
Model Predictive Control (MPC, see e.g. \cite{MRRS00,GoSD05}) is one of the few existing  techniques that is able to cope, in a quite straightforward way, with the presence of multiple inputs and outputs, of nonlinear dynamics and of hard constraints on the system state, $x$, and input, $u$. In MPC, at each time step $t$ the input $u(t)$ is computed by solving a finite horizon optimal control problem (FHOCP). The cost function to be minimized in the FHOCP is typically the average, over a finite horizon of $N<\infty$ steps, of the predicted values of a stage cost function, $l(x,u)$. The latter is chosen by the user, according to the goal to be achieved in the control problem at hand. In particular, there are two main classes of problems, giving rise to two different kinds of cost functions, respectively. In the first class, typically referred to as \emph{tracking MPC}, the aim is to drive the system state and input to reach a given set point or reference trajectory. The stage cost $l(x,u)$ employed in tracking problems is therefore related to the deviation of the predicted state and input trajectories from the reference ones. Most of the existing MPC formulations are concerned with this first class of problems, and a quite vast literature has been developed in the last decades \cite{MRRS00}, addressing nominal stability and recursive feasibility \cite{ScMR99}, as well as robustness analysis and robust design (see e.g. \cite{BeMo99b,GMTT04,GMTT05,MaRS06,LHRNB08}). In tracking MPC, the typical way to guarantee recursive feasibility of the FHOCP, as well as asymptotic stability of the target reference trajectory, is the use of a suitable cost function, of a sufficiently long horizon $N$ and/or of ``stabilizing constraints'', like state contraction constraints \cite{PoYa93,LoMo00}, Lyapunov-like constraints \cite{ScMR99}, terminal state constraints \cite{KeGi88} and terminal set constraints \cite{MiMa93}.\\
The second class of problems is that of \emph{economic MPC}, where the stage cost is not directly related to a prescribed set point or trajectory to be tracked, but it expresses a performance to be optimized. A typical application field, in which economic MPC is of high interest, is  process control, where the common approach consists of two hierarchical levels: at the upper level, a desired set point, according to the economic objective, is computed; at the lower level, a MPC law is used to track such a set point. In this context, economic MPC can be regarded as an integration of these two levels into a single predictive controller \cite{KaMa07,RaAm09}. More generally, economic MPC is an attractive approach for all control problems where the ``best'' performance, from the point of view of the economic objective, is not attained at any steady state or periodic trajectory, and/or one wants to avoid the pre-computation of a trajectory to be stabilized with tracking MPC. Economic MPC has been applied in practice in various fields, including process control \cite{KaMa07,Enge07,AsSS08}, renewable energy and energy efficiency \cite{HNID09,XiIl09,CaFM09c} and transportation \cite{KiMS09,VeBR10,BZVPKD10}, and the literature concerned with the theoretical properties of economic MPC schemes is all quite recent \cite{RaAm09,AnAR09,AnRw10,DiAR11,Grun11,AnAR11}. In most of the existing studies, a fixed point $(x^s,u^s)$ is computed that minimizes the average economic cost among all the admissible fixed points. Then, sufficient conditions on the FHOCP problem are derived, in order to make such a steady state asymptotically stable for the closed-loop system with an economic MPC law. In particular, in \cite{DiAR11,AnAR11} a terminal state constraint is used, to force the predicted state at time step $t+N$ to be equal to $x^s$, and conditions on the economic cost function are derived, under which asymptotic stability of $(x^s,u^s)$ is guaranteed. In \cite{AnAR11}, an asymptotic time-average economic  criterion is also introduced, in order to analyze the performance of economic MPC schemes. 
In \cite{Grun11}, the same time-average performance as \cite{AnAR11} is considered, but no terminal state constraint is used, and sufficient conditions on the prediction horizon and on the cost function are derived, under which the asymptotic time-average closed-loop performance is ``approximately optimal'', i.e. it converges to a value close to the minimal one.\\
In the described context, we investigate here the use of a terminal state constraint, which we call ``generalized'' because it requires the state at time step $t+N$ to be equal not to a specific fixed point, e.g. a set point to be tracked or a previously derived optimal fixed point, but to \emph{any} fixed point. This generalized terminal state constraint can be used for the design of either tracking or economic MPC schemes: in this paper, we study its properties in both contexts using a unified framework. In particular, we show that the use of the generalized terminal state constraint yields a larger feasibility set, with respect to a classical terminal state constraint approach. Moreover, we propose a novel receding horizon algorithm that, under mild assumptions, converges in finite time, with arbitrarily good accuracy, to an MPC law with an optimally chosen terminal state constraint, while still retaining a possibly larger feasibility set. Finally, we apply the approach to three examples. It has to be noted that this idea had been studied previously in the literature in the context of linear systems \cite{LAAC08,FLAAC09} and nonlinear ones \cite{FLAAC09b,FRLC10}, but with different assumptions and a different approach with respect to the one proposed here. A discussion about such differences is beyond the scope of this manuscript and will be included in future works. The paper is organized as follows. The problem settings are described in Section \ref{S:ProFor}; the generalized terminal state constraint, the related FHOCP, its receding horizon implementation and the recursive feasibility property are treated in Section \ref{S:New_constr}. Section \ref{S:New_MPC} is concerned with the guaranteed performance of the approach and the novel receding horizon implementation. Finally, examples are given in Section \ref{S:Example} and conclusions in Section \ref{S:Conclusions}.

\section{Notation and problem formulation}\label{S:ProFor}
We consider discrete-time  system models of the form:
\begin{equation}\label{E:system}
\begin{array}{l}
x(t+1)=f(x(t),u(t)),
\end{array}
\end{equation}
where $f:\R^n\times\R^m\rightarrow\R^n$, $t\in\mathbb{Z}$ is the discrete time variable, $x(t)\in\mathbb{R}^n$ is the system state and  $u(t)\in\mathbb{R}^m$ is the input.
State constraints are described by a set $\mathbb{X}\in\mathbb{R}^n$, and input constraints by a compact set $\mathbb{U}\in\mathbb{R}^m$. Mixed state-input constraints can be also considered, but they are omitted here for simplicity. The values of the generic variable $y$ at time $t+j$, predicted at time $t$, are indicated as $y(j|t),\,j\in\mathbb{N}$. Let $l:\R^n\times\R^m\rightarrow\R$ be a stage cost function, let $N\in\mathbb{N},\,0<N<\infty$ be a prediction horizon, finally define the cost function $J^s$ as:
\begin{equation}\label{E:cost_s}
J^s(x(t),U)\doteq\sum\limits_{j=0}^{N-1} l(x(j|t),u(j|t)),
\end{equation}
where $U=\{u(0|t),\ldots,u(N-1|t)\}$ is a sequence of $N$ predicted control inputs. Then, the following Finite Horizon Optimal Control Problem (FHOCP) $\cP^s(x(t))$ can be formulated:
\begin{subequations}\label{E:FHOCP}
\begin{align}&
\cP^s(x(t)):\nonumber\\
&\min\limits_{U}
J^s(x(t),U)\label{E:FHOCP_1}\\
&\text{subject to}\nonumber\\
&x(j|t)=f(x(j-1|t),u(j-1|t)),\,j=1,\ldots, N\label{E:FHOCP_2}\\
&u(j|t)\in\mathbb{U},\,\forall j=0,\ldots, N-1\label{E:FHOCP_3}\\
&x(j|t)\in\mathbb{X},\,\forall j=1,\ldots, N\label{E:FHOCP_4}\\
&x(0|t)=x(t)\label{E:FHOCP_5}\\
&x(N|t)=x^s\label{E:FHOCP_6},
\end{align}
\end{subequations}
where $x^s\in\mathbb{X}$ is fixed and chosen, together with the associated control input $u^s\in\mathbb{U}$, among the  (possibly multiple) fixed points $(x,u)$ that minimize the stage cost $l$ (see e.g. \cite{DiAR11}):
\begin{definition}\label{D:opt_fixed_point}(Optimal fixed point)
\begin{equation}\label{E:opt_steady_state}
\begin{array}{c}
(x^s,u^s)\in\arg\min\limits_{x\in\mathbb{X},u\in\mathbb{U}}l(x,u)\\
\text{subject to}\\
f(x,u)-x=0.
\end{array}
\end{equation}
\end{definition}
Problem $\cP^s$ is, in general, a nonlinear program (NLP) and, under mild regularity assumptions on $f$ and $l$, a (possibly local) minimum can be computed by using a numerical solver, indicated as $\lambda$. At a generic time step $t$, we denote such a solution as $U^*(x(t))$, and the corresponding optimal value as $J^{s*}(x(t))\doteq J^s(x(t),U^*(x(t)))$.

\begin{remark}
When $\cP^s$ is a general NLP, the outcome  $(J^{s*},U^*)$ of the numerical solver $\lambda$ is a function of both the parameter $x(t)$ and the starting sequence $\tilde{U}(x(t))$, with which the solver is initialized.
However, for the sake of simplicity of notation, we drop the dependence on $\tilde{U}(x(t))$, with the convention that, unless a starting sequence is explicitly specified, the solver $\lambda$ includes also an initialization procedure.
\end{remark}

\begin{remark}\label{R:convexity}
In many works on economic MPC, it is assumed that a global solution of $\cP^s(x(t))$ (as well as of problem \eqref{E:opt_steady_state}) can be computed, which is in general difficult to achieve for non-convex problems. Here, we first consider generally non-convex FHOCPs, and we will invoke convexity (or more generally the capability to always compute a global solution) only when necessary, in particular to prove performance and convergence results.
\end{remark}
The feasibility set $\cF^s$ is defined as follows:
\begin{definition}(Feasibility set)\\
$\cF^s\doteq\{x:\cP^s(x)\text{ admits a solution}\}$.
\end{definition}
Let $\cB(r,x)\doteq\{y:\|y-x\|_p\leq r\}$ for some $p\in[1,\infty)$. We consider the following assumption on the set $\cF^s$:
\begin{assumption}(Non-emptiness and boundedness of the feasibility set)\label{A:Feasible_set_1}
\begin{description}
  \item[I)] $\cF^s\neq \varnothing$
  \item[II)]$\exists r<\infty:\cF^s\subset\cB(r,0)$.
  \end{description}
\end{assumption}
Assumption \ref{A:Feasible_set_1} is quite general, since I) holds true if and only if the problem \eqref{E:opt_steady_state} is feasible, i.e. if there exists at least one fixed point that satisfies state and input constraints, and II) is either inherently satisfied by the FHOCP \eqref{E:FHOCP}, or it can be enforced in most practical applications, where typically the state values that are meaningful for the problem at hand are contained in a bounded set.\\
In MPC, the FHOCP \eqref{E:FHOCP} is solved at each time step by updating the measure of the state variable $x(t)$ according to a receding horizon (RH) strategy:\\
$\,$\\
\textbf{Algorithm 1} \emph{(RH control with terminal state constraint)
\begin{enumerate}
  \item (initialization) given $x(0)\in\cF^s$, let $t=0$, and solve the FHOCP $\cP^s(x(0))$; let $U^*(x(0))$   be a solution. Apply to the system the control input $u(0)=u^*(0|0)$. Set $t=1$;
  \item solve the FHOCP $\cP^s(x(t))$ by initializing the solver $\lambda$ with $\tilde{U}=\{u^*(1|t-1),\ldots,u^*(N-1|t-1),u^{s}\}$; let $U^*(x(t))$
  be a solution;
  \item apply to the system the control input $u(t)=u^*(0|t)$;
  \item  set $t=t+1$ and go to 2).
\end{enumerate}}
$\,$\\
We denote the state feedback control law, implicitly defined by Algorithm 1, as $u(t)=\kappa^s(x(t)),\,\kappa^s:\cF^s\rightarrow\mathbb{U}$. In the absence of noise and model uncertainty, for any given initial state $x(0)\in\cF^s$, Algorithm 1 guarantees recursive feasibility at all time steps $t>0$, i.e. $x(t)\in\cF^s,\,\forall t>0$. Recursive feasibility is achieved by initializing the solver, at step 2) of the algorithm, with the tail of the previous solution and the input $u^s$, corresponding to the steady state $x^s$. This, in turn, guarantees recursive satisfaction of state and input constraints, i.e. $x(t)\in\mathbb{X},\,\forall t>0,\,\kappa^s(x(t))\in\mathbb{U},\,\forall t\geq0$.\\
The stage cost $l(\cdot,\cdot)$ is chosen according to the considered control problem. In particular, in tracking MPC problems, the function $l(x,u)$ is often chosen as a quadratic function of the state and input tracking errors:
\begin{equation}\label{E:tracking_stage_cost}
l(x,u)=\|x-x^s\|_Q^2+\|u-u^s\|_R^2,
\end{equation}
where $\|y\|_M\doteq(y^TMy)^{1/2}$ and $Q=Q^\top$, $R=R^\top$, $Q,\,R\succ0$. With this choice (or, more generally, with any function $l$ such that $l(x,u)\geq0,\,\forall (x,u)\in\cF^s\times\mathbb{U},$ and such that $l(x,u)=0\iff (x,u)=(x^s,u^s)$), Algorithm 1 guarantees asymptotic convergence of the state and input trajectories to the optimal fixed point. With the addition of other, quite general assumptions on the regularity of $f$ and $l$ and with long enough horizon $N$, Algorithm 1 guarantees asymptotic stability of the fixed point $(x^s,u^s)$, with basin of attraction $\cF^s$ and some robustness margin (see e.g. \cite{MRRS00,GoSD05,GMTT04,GMTT05}).\\ In economic MPC problems, the stage cost $l$ is chosen according to some criterion that has to be minimized (or maximized), e.g. energy loss/production, fuel saving, etc.. In these cases, Algorithm 1 still guarantees recursive feasibility and state and input constraint satisfaction, however convergence and stability properties are not guaranteed in general, since they depend on the properties of the stage cost $l$. Sufficient conditions for asymptotic stability of the fixed point $(x^s,u^s)$ with an economic stage cost have been derived in \cite{DiAR11,AnAR11}. However, in economic MPC the stability of $(x^s,u^s)$ may be not relevant with respect to the control objective: in fact, while in tracking MPC the cost to be minimized attains its global minimum at the fixed point $(x^s,u^s)$, which can be regarded as the ``best'' operating point, in economic MPC the stage cost may not attain its minimum at any steady state, and a steady state solution might not be the most satisfactory operating condition for the system. In \cite{AnAR11}, an asymptotic time-average economic performance criterion, denoted here as $\overline{J}_\infty$, has been introduced, defined as:
\begin{equation}\label{E:average_economic_cost}
\overline{J}_\infty\doteq\lim\limits_{T\rightarrow\infty}\sup\frac{\sum\limits_{t=0}^T l(x(t),u(t))}{T+1}.
\end{equation}
The asymptotic average $\overline{J}_\infty$ appears to be more suited, with respect to stabilization of $(x^s,u^s)$, to represent the control objective in economic  MPC problems.
Clearly, in closed-loop operation the value of $\overline{J}_\infty$ is a function of the employed control law. In \cite{AnAR11}, it has been proved that:
\begin{equation}\label{E:average_economic_cost_kappa_s}
\overline{J}_\infty(\kappa^s)\leq l(x^s,u^s),
\end{equation}
thus showing that the use of Algorithm 1 gives an asymptotic time-average economic performance that is better than or equal to that of the stage cost associated to the ``best'' steady state.\\
In both tracking and economic MPC, the use of the FHOCP \eqref{E:FHOCP} in Algorithm 1 represents a straightforward way to achieve recursive feasibility and constraint satisfaction, however it is well known that the terminal state constraint \eqref{E:FHOCP_6} may be quite restrictive, so that typically quite ``long'' prediction horizons $N$ have to be employed to achieve a satisfactorily large feasibility set $\cF^s$, with a consequent higher computational complexity with respect to other techniques, like dual-mode MPC \cite{MRRS00}. In this paper, we adopt a particular terminal state constraint that aims to reduce this drawback, and we analyze the properties of the resulting closed-loop system in the case of both tracking and economic MPC.

\section{Generalized terminal state constraint}\label{S:New_constr}
Let $V=\{v(0|t),\ldots,v(N|t)\}\in\R^{m\times N+1}$ be a sequence of $N+1$ predicted control inputs, up to time $t+N$, let $\beta\in\R^+$ and $\overline{l}(t)\geq l(x^s,u^s)$ be two scalars, whose role will be better specified later on, and define the cost function $J$ as
\begin{equation}\label{E:cost_gen}
J(x(t),V)\doteq\sum\limits_{j=0}^{N-1}l(x(j|t),v(j|t))+\beta l(x(N|t),v(N|t)).
\end{equation}
Then, we propose to replace the FHOCP \eqref{E:FHOCP} with the following:
\begin{subequations}\label{E:gen_FHOCP}
\begin{align}
&\cP(\overline{l}(t),x(t)):\nonumber\\
&\min\limits_{V}
J(x(t),V)\label{E:gen_FHOCP_1}\\
&\text{subject to}\nonumber\\
 &x(j|t)=f(x(j-1|t),v(j-1|t)),\,j=1,\ldots,N\label{E:gen_FHOCP_2}\\
 &v(j|t)\in\mathbb{U},\,\forall j=0,\ldots,N\label{E:gen_FHOCP_3}\\
 &x(j|t)\in\mathbb{X},\,\forall j=1,\ldots,N\label{E:gen_FHOCP_4}\\
 &x(0|t)=x(t)\label{E:gen_FHOCP_5}\\
 &x(N|t)-f(x(N|t),v(N|t))=0\label{E:gen_FHOCP_6}\\
 &l(x(N|t),v(N|t))\leq\overline{l}(t)\label{E:gen_FHOCP_7}.
\end{align}
\end{subequations}
We denote a (possibly local) solution of $\cP(\overline{l}(t),x(t))$ as $V^*(\overline{l}(t),x(t))=\{v^*(0|t),\ldots,v^*(N|t)\}$, and the corresponding optimal value as $J^*(\overline{l}(t),x(t))\,\doteq J(x(t),V^*(\overline{l}(t),x(t)))$. Moreover, we indicate with $x^*(j|t),\,j\in[0,N]$ the sequence of predicted state values, computed by using the model \eqref{E:system}, starting from $x^*(0|t)=x(t)$ and applying the control sequence $V^*(\overline{l}(t),x(t))$.

\begin{remark}\label{R:variable_size}
The FHOCP $\cP(\overline{l}(t),x(t))$ has $(N+1)\,m$ optimization variables, i.e. $m$ more than problem $\cP^s(x(t))$ \eqref{E:FHOCP}. Depending on the considered application, this might or might not be an issue. This slight increase in the number of optimization variables can be seen as the ``price'' for generalizing the terminal state constraint \eqref{E:gen_FHOCP_6}. Moreover, we show in our examples how the use of a generalized terminal state constraint, with a much shorter horizon $N$, can yield closed-loop performance that are similar to those obtained with a fixed terminal state constraint and longer horizon, thus effectively reducing the computational effort.
\end{remark}

The generalized feasibility set $\cF$ is defined as:

\begin{definition}(Generalized feasibility set)\label{D:gen_feasibility_set}\\
$\cF\doteq\{x:\cP(\overline{l},x)\text{ admits a solution for some }\overline{l}\}$.
\end{definition}

For a given $x(t)\in\cF$, let us define the set $\cX(x(t),N)$ as follows:
\begin{definition}\label{D:reachable_fixed}(Set of reachable fixed points)
\begin{equation}\label{E:reachable_fixed_points}
\begin{array}{l}
\cX(x(t),N)\doteq\\\{x\in\X:
\exists V\in\R^{m\times N+1}:v(j|t)\in\U,\,\forall j\in[0,N];\\
x(N|t)=x;\,f(x,v(N|t))=x;\\
x(j|t)=f(x(j-1|t),v(j-1|t)),\,\forall j\in[1,N]\\
x(j|t)\in\X,\,\forall j\in[1,N]\}.
\end{array}
\end{equation}
\end{definition}
The set $\cX(x(t),N)$ contains all the possible steady state values that can be reached in at most $N$ steps with an admissible control sequence $V$, starting from $x(t)$. It is straightforward to note that if $N^1>N^2$, then $\cX(x(t),N^1)\supseteq\cX(x(t),N^2)$. The following result is concerned with the relationship between the sets $\cX(x(t),N)$ and $\cF^s$.
\begin{proposition}\label{P:reachable_set_feasibility_set}
Let Assumption \ref{A:Feasible_set_1} hold. Then:
\begin{equation}\label{E:reach_feas_rel}
x^s\in\cX(x(t),N)\iff x(t)\in\cF^s
\end{equation}
\end{proposition}
\begin{pf}
If $x^s\in\cX(x(t),N)$, then by how the set $\cX(x(t),N)$ is defined, there exists a sequence $V$ of $N+1$ predicted control moves, such that all of the constraints \eqref{E:gen_FHOCP_2}-\eqref{E:gen_FHOCP_6} are satisfied, with $x(N|t)=x^s$. Therefore, the first $N$ elements of such a sequence  satisfy also constraints \eqref{E:FHOCP_1}-\eqref{E:FHOCP_6}, hence $x(t)\in\cF^s$. Conversely, if $x(t)\in\cF^s$, then there exists an optimal solution $U^*(x(t))$ satisfying constraints \eqref{E:FHOCP_1}-\eqref{E:FHOCP_6}. Thus, the sequence $V=\{U^*(x(t)),\,u^*(N|t)\}$ satisfies the constraints in \eqref{E:reachable_fixed_points} with $x(N|t)=x^s$, i.e. $x^s\in\cX(x(t),N)$.\hfill$\Box$
\end{pf}
We also define the quantity $\underline{l}(x(t))$ as:
\begin{definition}\label{D:opt_achievable_cost}(Optimal achievable stage cost)
For a given $x(t)\in\cF$, the optimal achievable stage cost is:
\begin{equation}\label{E:optimal_stage_t}
\begin{array}{ccc}
\underline{l}(x(t))&\doteq&\min\limits_{x\in\cX(x(t),N),u\in\U}l(x,u)\\
& &\text{subject to}\\
& & f(x,u)=x.
\end{array}
\end{equation}
\end{definition}

\begin{assumption}\label{A:opt_achiev_cost_exists}(Existence of the optimal achievable cost)
For any $x\in\cF$, the value $\underline{l}(x(t))$ of Definition \ref{D:opt_achievable_cost} exists.
\end{assumption}

Assumption \ref{A:opt_achiev_cost_exists} holds in most practical cases, considering that the input constraint set $\U$ is compact, the horizon $N$ is finite and the stage cost $l$ can be chosen by the designer.\\
We can now define the set $\mathcal{S}\doteq\{(\overline{l},x):\overline{l}\geq \underline{l}(x),\,x\in\cF\}$, 
as well as the following functions:
\begin{equation}\label{E:functions_implicit}
\begin{array}{l}
\kappa(\overline{l}(t),x(t))\doteq v^*(0|t)\\
\zeta(\overline{l}(t),x(t))\doteq l(x^*(N|t),v^*(N|t))\\
\kappa:\mathcal{S}\rightarrow\U\\
\zeta:\mathcal{S}\rightarrow\mathcal{S}
\end{array}
\end{equation}
The value $\kappa(\overline{l}(t),x(t))$ corresponds to the first control input in the sequence $V^*(\overline{l}(t),x(t))$ and the value $\zeta(\overline{l}(t),x(t))$ is the cost associated to the terminal state-input pair, obtained by applying to system \eqref{E:system} the sequence $V^*(\overline{l}(t),x(t))$, starting from the initial condition $x(t)$.
$\,$\\
The following RH strategy is obtained by recursively solving the FHOCP \eqref{E:gen_FHOCP}:\\
$\,$\\
\textbf{Algorithm 2} \emph{(RH control with generalized terminal state constraint)
\begin{enumerate}
  \item (initialization) choose a value of $\beta>0$. Given $x(0)\in\cF$, choose a value $\overline{l}(0)$ such that $(\overline{l}(0),x(0))\in\mathcal{S}$ and let $t=0$. Solve the FHOCP $\cP(\overline{l}(0),x(0))$; let $V^*(\overline{l}(0),x(0))$  be a solution. Apply to the system the control input $u(0)=\kappa(\overline{l}(0),x(0))$. Set $t=1$;
  \item set $\overline{l}(t)=\zeta(\overline{l}(t-1),x(t-1))$ and solve the FHOCP $\cP(\overline{l}(t),x(t))$ by initializing the solver $\lambda$ with $\tilde{V}=\{v^*(1|t-1),\ldots,v^*(N|t-1),v^*(N|t-1)\}$; let $V^*(\overline{l}(t),x(t))$
  be a solution;
  \item apply to the system the control input $u(t)=\kappa(\overline{l}(t),x(t))$;
  \item  set $t=t+1$ and go to 2).
\end{enumerate}}

\begin{remark}\label{R:Algorithm2} From a practical point of view, we note that the value of $\underline{l}(x(0))$ needs not to be known explicitly in the initialization step of Algorithm 2, when selecting $\overline{l}(0)\in\mathcal{S}$: in fact, by construction any value of $\overline{l}(0)$ such that the problem $\cP(\overline{l}(0),x(0))$ is feasible belongs to $\mathcal{S}$.\end{remark}

The application of Algorithm 2 gives rise to the following closed-loop system: 
\begin{subequations}\label{E:gen_closed_loop}
\begin{align}
&x(t+1)=f(x(t),\kappa(\overline{l}(t),x(t)))\label{E:gen_closed_loop_1}\\
&\overline{l}(t+1)=\zeta(\overline{l}(t),x(t))\label{E:gen_closed_loop_2}
\end{align}
\end{subequations}
We denote with $\psi(k,\overline{l}(t),x(t))$ and $\phi(k,\overline{l}(t),x(t))$ the values of the bound $\overline{l}(t+k)$ and of the state $x(t+k)$, respectively, at the generic time $t+k,\,k\in\mathbb{N},$ obtained by applying \eqref{E:gen_closed_loop} starting from $x(t)$ and $\overline{l}(t)$.\\

Our first result is concerned with the existence of $\cF$ and its relationship with the set $\cF^s$ and with the properties of recursive feasibility of problem $\cP(\overline{l}(t),x(t))$ in Algorithm 2, hence of the capability of the control law $\kappa$ to satisfy input and state constraints.

\begin{thm}\label{T:gen_feasibility}
Let Assumption \ref{A:Feasible_set_1} hold, and consider the closed-loop system \eqref{E:gen_closed_loop}, obtained by applying Algorithm 2 with any $\beta\geq0$ in the FHOCP $\cP$. The following properties hold:
\begin{description}
  \item[\textbf{a)}] (feasibility set)\\
  $\cF\supseteq\cF^s$.
  \item[\textbf{b)}] (recursive feasibility)\\
  $\cP(\psi(t,\overline{l}(0),x(0)),\phi(t,\overline{l}(0),x(0)))$ is feasible\\ $\forall (\overline{l}(0),x(0))\in\mathcal{S},\,\forall t>0$
  \item[\textbf{c)}] (state constraint satisfaction)\\ $\phi(t,\overline{l}(0),x(0))\in\mathbb{X},\,\forall (\overline{l}(0),x(0))\in\mathcal{S},\,\forall t>0$
  \item[\textbf{d)}] (input constraint satisfaction)\\ $\kappa(\psi(t,\overline{l}(0),x(0)),\phi(t,\overline{l}(0),x(0)))\in\mathbb{U},\,\forall (\overline{l}(0),x(0))\in\mathcal{S},\,\forall t\geq0$.
\end{description}
\end{thm}

\begin{pf}\textbf{a)} By Assumption \ref{A:Feasible_set_1}, there exists a set $\cF^s$ of state values such that problem $\cP^s$ is feasible. Then, it is straightforward to note that also problem $\cP(l(x^s,u^s),x(t))$ is feasible for all $x(t)\in\cF^s$.\\
\textbf{b)} For any given $(\overline{l}(0),x(0))\in\mathcal{S}$, by definition problem $\cP(\overline{l}(0),x(0))$ admits a solution $V^*(\overline{l}(0),x(0))$.  Clearly, the pair $(x^*(N|0),v^*(N|0))$ is  a fixed point for system \eqref{E:system}. According to Algorithm 2 the control move $u(0)=v^*(0|0)$ is applied. At $t=1$, the new state value is $x(1)=\phi(1,\overline{l}(0),x(0))=f(x(0),v^*(0|t))=x^*(1|t)$, and the solver $\lambda$ is initialized with the starting sequence $\tilde{V}$, which includes the tail $\{v^*(1|0),\ldots,v^*(N|0)\}$ of sequence $V^*(\overline{l}(0),x(0))$, plus the terminal input $\tilde{v}(N|1)=v^*(N|0)$. Let us denote  with $\tilde{x}(j|1),\,j\in[0,N]$, the corresponding sequence of predicted state values. Then, it can be noted that $\tilde{x}(N-1|1)=\tilde{x}(N|1)=x^*(N|0)$, i.e. the pair $(\tilde{x}(N|1),\tilde{v}(N|1))=(x^*(N|0),v^*(N|0))$ satisfies constraint \eqref{E:gen_FHOCP_6}. Constraints \eqref{E:gen_FHOCP_2}-\eqref{E:gen_FHOCP_4} are also satisfied, since the predicted state and input sequences are the same as the ones computed at the previous time step, plus an additional state-input pair $(\tilde{x}(N|1),\tilde{v}(N|1))$. Finally, constraint \eqref{E:gen_FHOCP_7} is satisfied too, since $\psi(1,\overline{l}(0),x(0))=\overline{l}(1)=\zeta(\overline{l}(0),x(0))=l(x^*(N|0),v^*(N|0))=l(\tilde{x}(N|1),\tilde{v}(N|1))$.  Therefore, the sequence $\tilde{V}$ is admissible for problem\\ $\cP(\psi(1,\overline{l}(0),x(0)),\phi(1,\overline{l}(0),x(0)))$. The same reasoning applies recursively for all $t\geq0$.\\
\textbf{c)-d)} Straightforward consequences of the recursive feasibility property.\hfill$\Box$
\end{pf}

\begin{remark}\label{R:feasibility}
According to Theorem \ref{T:gen_feasibility}, the generalized feasibility set $\cF$ is positively invariant for the trajectories $\phi(t,\overline{l}(0),x(0))$ of system \eqref{E:system} with the feedback control law $\kappa$, and it is at least as large as the feasibility set of problem $\cP^s$. This means that, in the ``worst'' case, for given $x(0)$ problem $\cP$ is feasible if and only if $\cP^s$ is, but, depending on the system \eqref{E:system}, it may happen that $\cP$ is feasible for a larger set of state values, given the same prediction horizon $N$, or alternatively that $\cP$ is similar to $\cP^s$, but with a shorter prediction horizon, i.e. lower computational complexity. Moreover, the set $\mathcal{S}$ is also forward invariant for the values of $(\psi(t,\overline{l}(0),x(0)),\phi(t,\overline{l}(0),x(0)))$.\end{remark}

The generalized feasibility set may be larger than the feasibility set obtained with a fixed terminal state constraint; however nothing can be said, in general, about the performance of the closed-loop system obtained by applying Algorithm 2, in terms of stability of the target steady state in tracking MPC, and of asymptotic average cost in economic MPC. In fact, the performance guarantees achieved by Algorithm 1, with a fixed terminal state constraint, are a direct consequence of the fact that the employed value of $(x^s,u^s)$ has been optimally chosen off-line, according to the control problem at hand. On the contrary, in Algorithm 2, the terminal state and input $(x^*(N|t),v^*(N|t))$ are different, in general, from the values $(x^s,u^s)$, and they are allowed to change at each time step $t$. Basically, the values of $(x^*_{N|t},v^*_{N|t})$ are implicitly ``selected'', among all the possible steady states that can be reached in at most $N$ steps from the actual state $x(t)$, by the numerical solver $\lambda$, in order to minimize the cost $J(x,V)$ \eqref{E:cost_gen}. Therefore, for given control horizon $N$ and constraints $\X,\,\U$, the values $(x^*(N|t),v^*(N|t))$ depend on the chosen stage cost $l(\cdot,\cdot)$ and on the scalar weight $\beta$. Moreover, it can be noted that the use of Algorithm 2 gives rise to a sequence of pairs $\{(x^*(N|t),v^*(N|t))\}_{t=0}^\infty$, and consequently a sequence of terminal cost values $\{l(x^*(N|t),v^*(N|t))\}_{t=0}^\infty$. The performance achieved by the system \eqref{E:gen_closed_loop} clearly depends on the behavior of such a sequence. In this regard, we note that the control law $u(t)=\kappa(\overline{l}(t),x(t))$, obtained by using Algorithm 2, is the output of a dynamical system, with internal state $\overline{l}(t)$ and input $x(t)$. This is in contrast with the typical MPC control laws, like $\kappa^s$, that are static feedback controllers. The controller's state $\overline{l}(t)$ traces the value of the stage cost associated with the terminal state-input pair $l(x^*(N|t),v^*(N|t))$, hence it carries the information about how suboptimal is such a terminal cost with respect to the optimal one, $l(x^s,u^s)$. The inequality $l(x^*(N|t),v^*(N|t))\leq l(x^*_{N|t-1},v^*_{N|t-1})=\overline{l}(t)$, enforced by means of constraint \eqref{E:gen_FHOCP_7}, ensures that  the sequence $\{l(x^*_{N|t},v^*_{N|t})\}_{t=0}^\infty$ is not increasing, however,  in the general settings considered so far, there is no guarantee of convergence to the optimal value $l(x^s,u^s)$, or to a value close to the optimal. As a consequence, no guaranteed performance properties can be obtained. One option to deal with this issue is to use Algorithm 2 as it is, to set some initial choices of $N$, $l$ and $\beta$ and to tune these parameters following a trial-and-error procedure, in order to obtain a satisfactory closed-loop performance. Indeed, quite good results can be typically obtained in this way, as highlighted in the examples of Section \ref{S:Example}. Another option is to consider additional assumptions on the problem, to derive guidelines on how to choose $N$, $l$ and $\beta$, as well as to adopt a more sophisticated receding horizon algorithm, in order to guarantee a desired behavior of the sequence $\{l(x^*(N|t),v^*(N|t))\}_{t=0}^\infty$, in terms of convergence to a value which is arbitrarily close to the optimal one, $l(x^s,u^s)$. Such a modified algorithm and its properties are described in the next section.

\section{Guaranteed properties of MPC with generalized terminal state constraint}\label{S:New_MPC}

We first establish sufficient conditions on $\beta$ under which the terminal state and input pair $(x^*(N|t),v^*(N|t))$, computed by solving problem $\cP(\overline{l}(t),x(t))$, has an associated cost $l(x^*(N|t),v^*(N|t))$ which is arbitrarily close to the minimal one, among all the possible steady states that can be reached from $x(t)$. In order to do so, we consider the next three assumptions. We recall that a continuous, monotonically increasing function $\alpha:[0,+\infty)\rightarrow[0,+\infty)$ is a $\cK_\infty$ function if $\alpha(0)=0$ and $\lim\limits_{a\rightarrow+\infty}\alpha(a)=+\infty$, and denote such functions as $\alpha\in\cK_\infty$.

\begin{assumption}\label{A:fl_continuity}(Boundedness of the generalized feasibility set and continuity of $f$ and $l$)
\begin{description}
  \item[I)]$\exists r<\infty:\cF\subset\cB(r,0)$;
  \item[II)] $f$ and $l$ are continuous on $\overline{\cF}\times\U$, where $\overline{\cF}$ is the closure of $\cF$, hence $\exists\, \alpha_f,\alpha_l\in\cK_\infty:$$\|f(\bar{x},\bar{u})-f(\hat{x},\hat{u})\|\leq\alpha_f(\|(\bar{x},\bar{u})-(\hat{x},\hat{u})\|),\,$
$|l(\bar{x},\bar{u})-l(\hat{x},\hat{u})|\leq\alpha_l(\|(\bar{x},\bar{u})-(\hat{x},\hat{u})\|),$
$\forall(\bar{x},\bar{u}),(\hat{x},\hat{u})\in\cF\times\U$, for some vector norm $\|\cdot\|$.
  \end{description}
\end{assumption}

Similarly to the set $\cF^s$, the set $\cF$ in many cases is bounded in the presence of bounded state and input constraints, while, depending on the system equations $f$, it might be unbounded if the state constraint set $\mathbb{X}$ is unbounded, even with a finite horizon $N$, due to the generalized terminal state constraint \eqref{E:gen_FHOCP_6}. However, in practical applications the  initial state values that are meaningful for the problem at hand are typically contained in a bounded set $\mathbb{F}\subset\R^n$, so that one can always consider a ``reduced'' feasibility set $\tilde{\cF}=\cF\cap\mathbb{F}$ to satisfy Assumption \ref{A:fl_continuity}-I).

\begin{assumption}\label{A:solution_opt} (Solution of the FHOCP $\cP$)\\
For any $(\overline{l},x)\in\mathcal{S}$, the FHOCP $\cP(\overline{l},x)$ has at least one global minimum, which is computed by the solver $\lambda$ independently on how it is initialized.
\end{assumption}

As already discussed in Remark \ref{R:convexity}, Assumption \ref{A:solution_opt} is quite usual in the context of economic MPC. Moreover, it is satisfied if problem $\cP(\overline{l},x)$ is convex, which is the important case of MPC for linear systems with convex constraints $\X,\,\U$ and convex stage cost $l$.

\begin{assumption}\label{A:min_l} (Stage cost)\\
There exists a set $\mathcal{M}\subset\cF\times\mathbb{U}$ where the function $l$ attains its minimum. Without loss of generality, $l(x,u)\geq0,\,\forall(x,u)\in\cF\times\mathbb{U}$, and $l(x,u)=0\iff(x,u)\in\mathcal{M}$.
\end{assumption}

Assumption \ref{A:min_l} is obviously satisfied for tracking MPC with stage costs like \eqref{E:tracking_stage_cost}, with $\mathcal{M}=\{(x^s,u^s)\}$. In the case of economic MPC, satisfaction of Assumption \ref{A:min_l} depends on the stage cost chosen by the control designer, and the set  $\mathcal{M}$ typically does not contain any steady state and might also be not connected.\\

We can now derive a result related to the optimality of the pair $(x^*(N|t),v^*(N|t))$ with respect to the stage cost function $l$.
\begin{proposition}\label{P:opt_terminal}
Let Assumptions \ref{A:Feasible_set_1}-\ref{A:solution_opt} hold. Then, for any $\epsilon>0$, there exists a finite value $\underline{\beta}(\epsilon)$ such that, for any given $x(t)\in\cF$ and any $\overline{l}(t)\geq\underline{l}(x(t))+\epsilon$, if $\beta\geq\underline{\beta}(\epsilon)$ then
\begin{equation}\label{E:optimal_prediction}
l(x^*(N|t),v^*(N|t))\leq\underline{l}(x(t))+\epsilon
\end{equation}
where $(x^*(N|t),v^*(N|t))$ are the optimal terminal state and input computed by solving problem $\cP(\overline{l}(t),x(t))$.
\end{proposition}
\begin{pf}
Let $(\overline{x},\overline{u})$ be a state-input pair such that:
\[
\begin{array}{ccc}
(\overline{x},\overline{u})&=&\arg\min\limits_{x\in\cX(x(t),N),u\in\U}l(x,u)\\
& &\text{subject to}\\
& & f(x,u)=x,
\end{array}
\]
and let $\overline{V}$ be a sequence of $N+1$ control inputs which satisfies constraints \eqref{E:gen_FHOCP_2}-\eqref{E:gen_FHOCP_6} and such that $x(N|t)=\overline{x},\,v(N|t)=\overline{u}$. This sequence is guaranteed to exist by Definition \ref{D:reachable_fixed}. Moreover, we have $l(\overline{x},\overline{u})=\underline{l}(x(t))<\overline{l}(t)$, hence also constraint \eqref{E:gen_FHOCP_7} is satisfied and the sequence $\overline{V}$ is admissible for problem $\cP(\overline{l}(t),x(t))$. The cost associated to $\overline{V}$ is equal to:
\[
J(x(t),\overline{V})=\sum\limits_{j=0}^{N-1}l(\overline{x}(j|t),\overline{v}(j|t))+\beta \underline{l}(x(t)),
\]
where $\overline{x}(j|t),\,j\in[0,N]$ is the state trajectory obtained by applying the sequence $\overline{V}$.
Consider now any other possible state $\hat{x}\in\cX(x(t),N)$ and input $\hat{u}\in\U$ such that $f(\hat{x},\hat{u})=\hat{x}$, $l(\hat{x},\hat{u})>\underline{l}(x(t))+\epsilon$ and $l(\hat{x},\hat{u})\leq\overline{l}(t)$. Denote with $\hat{V}$ a sequence of control inputs which is feasible for problem $\cP(\overline{l}(t),x(t))$ and such that $x(N|t)=\hat{x}$. Then, the cost associated to $\hat{V}$ is:
\[
J(x(t),\hat{V})=\sum\limits_{j=0}^{N-1}l(\hat{x}(j|t),\hat{v}(j|t))+\beta l(\hat{x},\hat{u}),
\]
where $\hat{x}(j|t),\,j\in[0,N]$ is the state trajectory obtained by applying the sequence $\hat{V}$.
Thus, the difference $J(x(t),\overline{V})-J(x(t),\hat{V})$ is given by:
\begin{equation}\label{E:proof_prop_aux}
\begin{array}{l}
J(x(t),\overline{V})-J(x(t),\hat{V})=\beta[\underline{l}(x(t))-l(\hat{x},\hat{u})]+\\
\sum\limits_{j=0}^{N-1}[l(\overline{x}(j|t),\overline{v}(j|t))-l(\hat{x}(j|t),\hat{v}(j|t))],
\end{array}
\end{equation}
and, by exploiting Assumption \ref{A:fl_continuity}, it holds:
\begin{equation}\label{E:bound_cost_diff}
J(x(t),\overline{V})-J(x(t),\hat{V})<-\beta\epsilon+\eta,
\end{equation}
with (see the Appendix for a complete proof)
\begin{equation}\label{E:eta}
\eta=\alpha_l\left(\sum\limits_{i=0}^{N-1}\alpha_f^{(i-j)}\left(\max\limits_{\overline{v},\hat{v}\in\U}\|\overline{v}-\hat{v}\|\right)\right)>0,
\end{equation}
where $\alpha_f^{(i)}(a)\doteq\underbrace{\alpha_f(\alpha_f(\ldots\alpha_f(a)\ldots))}\limits_{\text{$i$ times}}$ and $\alpha^{(0)}(a)\doteq a$.
Thus, by setting $\underline{\beta}(\epsilon)=\eta/\epsilon$ and $\beta\geq\underline{\beta}(\epsilon)$, the inequality
\begin{equation}\label{E:opt_cost_proof}
J(x(t),\hat{V})> J(x(t),\overline{V})
\end{equation}
is obtained. Note that, due to the compactness of $\U$, the value of $\eta$ is finite, hence also $\underline{\beta}(\epsilon)$ is finite.
Now assume, with the purpose of contradiction, that $l(x^*(N|t),v^*(N|t))>\underline{l}(x(t))+\epsilon$. Then,  inequality \eqref{E:opt_cost_proof} would hold true with $\hat{V}=V^*(\overline{l}(t),x(t))$, meaning that the cost associated to $V^*(\overline{l}(t),x(t))$ would be higher than the one associated to $\overline{V}$. However, this cannot happen, since, by Assumption \ref{A:solution_opt}, $V^*(\overline{l}(t),x(t))$ is such that $J(x(t),V^*(\overline{l}(t),x(t)))=J^*(\overline{l}(t),x(t))\leq J(x(t),V)$ for any $V$ which is feasible for problem $\cP(\overline{l}(t),x(t))$. Hence, the inequality \eqref{E:optimal_prediction} must hold.\hfill$\Box$
\end{pf}
Proposition \ref{P:opt_terminal} provides an indication on how to tune the parameter $\beta$ in \eqref{E:gen_FHOCP}: the higher this value, the closer the terminal stage cost is to the best one achievable starting from the current state $x(t)$. This property induces a result pertaining to the performance achieved by using Algorithm 2 when the initial state $x(0)$ belongs to the feasibility set $\cF^s$. Before stating such a result, we consider the following assumption for tracking MPC schemes.

\begin{assumption}\label{A:tracking_stage_cost}(Stage cost in tracking MPC)\\
In tracking MPC, the stage cost $l$ enjoys the following properties:
\begin{description}
  \item[I)](global minimum)\\
    \begin{equation}\label{E:optimal_stage_cost_tracking}
    \begin{array}{l}
    l(x,u)>0,\,\forall (x,u)\in\cF^s\times\U\setminus\{(x^s,u^s)\}\\
    l(x^s,u^s)=0
    \end{array}
    \end{equation}
  \item[II)] (lower bound)\\
    \begin{equation}\label{E:l_lower_bound}
    \begin{array}{l}
    \exists\underline{\alpha}_l\in\cK_\infty:\\
    \underline{\alpha}_l(\|(x,u)-(x^s,u^s)\|)\leq l(x,u),\,\forall (x,u)\in\cF^s\times\U.
    \end{array}
    \end{equation}
\end{description}
\end{assumption}

Note that Assumption \ref{A:tracking_stage_cost} is typically satisfied by the stage cost functions used in tracking MPC, like \eqref{E:tracking_stage_cost}.

\begin{thm}\label{T:perf_feas_set_s}
Let Assumptions \ref{A:Feasible_set_1}-\ref{A:min_l} hold, let a value of $\epsilon>0$ be chosen, and let $\beta\geq\underline{\beta}(\epsilon)$. For any $x(0)\in\cF^s$, apply Algorithm 2. Then, the following properties hold:
\begin{description}
  \item[\textbf{a)}](sub-optimality of the terminal stage cost)\\
    \begin{equation}\label{E:terminal_state_cost_diff}
    l(x^*(N|t),v^*(N|t))-l(x^s,u^s)\leq\epsilon,\,\forall t\geq0,
    \end{equation}
  \item[\textbf{b)}] (tracking MPC) if Assumption \ref{A:tracking_stage_cost} also holds, then:
\begin{equation}\label{E:terminal_state_opt}
\|(x^*(N|t),v^*(N|t))-(x^s,u^s)\|\leq\underline{\alpha}_l^{-1}(\epsilon),\,\forall t\geq0,
\end{equation}
  \item[\textbf{c)}] (economic MPC) the asymptotic average performance obtained by control law $\kappa$ is bounded as: \\ \begin{equation}\label{E:average_economic_cost_kappa}
        \overline{J}_\infty(\kappa)\leq l(x^s,u^s)+\epsilon.
        \end{equation}

\end{description}
\end{thm}

\begin{pf}
\textbf{a)} According to Proposition \ref{P:reachable_set_feasibility_set}, if $x(0)\in\cF^s$ then $x^s\in\cX(x(0),N)$. Moreover, by Definitions \ref{D:opt_fixed_point} and \ref{D:opt_achievable_cost}, if $x^s\in\cX(x(0),N)$ then $\underline{l}(x(0))=l(x^s,u^s)$. Therefore, by Proposition \ref{P:opt_terminal} we have $l(x^*(N|0),v^*(N|0))-l(x^s,u^s)\leq\epsilon$. The use of Algorithm 2 and constraint \eqref{E:gen_FHOCP_7} force the values $l(x^*(N|t),v^*(N|t))$ to be not increasing with $t$, thus the inequality $l(x^*(N|t),v^*(N|t))-l(x^s,u^s)\leq\epsilon$ holds true for all $t\geq0$.\\
\textbf{b)} From \eqref{E:terminal_state_cost_diff}, under Assumption \ref{A:tracking_stage_cost}-I) we have
\[\begin{array}{l}
l(x^*(N|t),v^*(N|t))-\,l(x^s,u^s)\\
=l(x^*(N|t),v^*(N|t))\leq\epsilon,\,\forall t\geq0.
\end{array}
\]
Then, by Assumption \ref{A:tracking_stage_cost}-II) it holds $\|(x^*(N|t),v^*(N|t))-(x^s,u^s)\|\leq\underline{\alpha}_l^{-1}(\epsilon),\,\forall t\geq0$.\\
\textbf{c)} The proof of this claim follows that of Theorem 1 in \cite{AnAR11}, with little modifications, and it is reported here for the sake of completeness. First of all, note that, for any $t\geq0$, it holds:
\[
\begin{array}{l}
J^*(\overline{l}(t),x(t))=l(x(t),u(t))+\sum\limits_{j=1}^{N-1}l(x^*(j|t),v^*(j|t))+\\
\beta l(x^*(N|t),v^*(N|t));\\
J^*(\overline{l}(t+1),x(t+1))\leq \sum\limits_{j=1}^{N-1}l(x^*(j|t),v^*(j|t))\\+l(x^*(N|t),v^*(N|t))+\beta l(x^*(N|t),v^*(N|t)),
\end{array}
\]
thus, under Assumptions \ref{A:Feasible_set_1}-\ref{A:min_l}, by using \eqref{E:terminal_state_cost_diff}, for any $x(0)\in\cF^s$ it holds:
\[
\begin{array}{rl}
&J^*(\overline{l}(t+1),x(t+1))-J^*(\overline{l}(t),x(t))\\ &\leq l(x^s,u^s)+\epsilon-l(x(t),u(t))\\
\Rightarrow&\lim\limits_{T\rightarrow\infty}\inf\frac{\sum\limits_{t=0}^TJ^*(\overline{l}(t+1),x(t+1))-J^*(\overline{l}(t),x(t))}{T+1}\\
\leq&\lim\limits_{T\rightarrow\infty}\inf\frac{\sum\limits_{t=0}^Tl(x^s,u^s)+\epsilon-l(x(t),u(t))}{T+1}\\
=& l(x^s,u^s)+\epsilon-\lim\limits_{T\rightarrow\infty}\sup\frac{\sum\limits_{t=0}^T l(x(t),u(t))}{T+1}.
\end{array}
\]
At the same time, due to Assumption \ref{A:min_l},
\[
\begin{array}{rl}
&\lim\limits_{T\rightarrow\infty}\inf\frac{\sum\limits_{t=0}^TJ^*(\overline{l}(t+1),x(t+1))-J^*(\overline{l}(t),x(t))}{T+1}\\
=&\lim\limits_{T\rightarrow\infty}\inf\frac{J^*(\overline{l}(T+1),x(T+1))-J^*(\overline{l}(0),x(0))}{T+1}\\
\geq&\lim\limits_{T\rightarrow\infty}\inf\frac{-J^*(\overline{l}(0),x(0))}{T+1}=0\\
\end{array}
\]
hence
\[
\lim\limits_{T\rightarrow\infty}\sup\frac{\sum\limits_{t=0}^Tl(x(t),u(t))}{T+1}=\overline{J}_\infty(\kappa)\leq l(x^s,u^s)+\epsilon
\]
This bound establishes the result.\hfill$\Box$
\end{pf}
According to Theorem \ref{T:perf_feas_set_s}-\textbf{a)}, with a sufficiently high value of $\beta$, for any $x(0)$ inside the feasibility set of problem $\cP^s$, the generalized terminal stage cost is always at most $\epsilon$-suboptimal with respect to the one related to the optimal pair $(x^s,u^s)$ \eqref{E:opt_steady_state}. Theorem \ref{T:perf_feas_set_s}-\textbf{b)} implies that it is possible to force the generalized terminal state-input pair to be arbitrarily close, as $\epsilon\rightarrow 0$, to the desired one, for any $x(0)\in\cF^s$.  As a consequence, the convergence and stability properties of tracking MPC schemes, with a fixed terminal state constraint, can be extended to the case of generalized terminal state constraint, by considering an arbitrarily small neighborhood of the desired set point $(x^s,u^s)$. Finally, Theorem \ref{T:perf_feas_set_s}-\textbf{c)} states that the MPC scheme with generalized terminal state constraint achieves an asymptotic average performance which is better than that of the optimal fixed point $(x^s,u^s)$, plus an arbitrarily small tolerance $\epsilon$.

\begin{remark}\label{R:optimal_terminal}
Proposition \ref{P:opt_terminal} provides only a sufficient condition on $\beta$, thus it is often conservative: as a matter of fact, with a reasonably high value of $\beta$ the performance of the MPC scheme with generalized terminal state constraint, described in Theorem \ref{T:perf_feas_set_s}, matches those obtained with a fixed, optimally chosen terminal state constraint over all the set $\cF^s$. This aspect is highlighted in the examples of Section \ref{S:Example}.
\end{remark}

Practically speaking, Theorem \ref{T:perf_feas_set_s} states that the control law $\kappa(x)$ can achieve performance and stability properties that are arbitrarily close to those of $\kappa^s(x)$, for any $x(0)\in\cF^s$. For state values $x(0)\in\cF\setminus\cF^s$, such a comparison does not make sense, since the control law $\kappa^s$ is not defined outside the set $\cF^s$, and the optimal fixed point $x^s$ is not reachable (see Proposition \ref{P:reachable_set_feasibility_set}). We now focus our attention on initial state values $x(0)\in\cF\setminus\cF^s$, and we present a modified algorithm that guarantees that the resulting MPC law enjoys the properties of Theorem \ref{T:perf_feas_set_s}, under the following additional assumption.
\begin{assumption}\label{A:steady_state_sequence} (Sequences of steady states with decreasing stage cost)\\
For some (eventually very small) $\overline{\epsilon}>0$ there exists a minimal prediction horizon $\underline{N}\in\mathbb{N}$ such that, for any fixed point $(\overline{x},\overline{u})\in\cF\times\U:\,f(\overline{x},\overline{u})=\overline{x}$, there exists at least one state-input pair $(\tilde{x},\tilde{u})$ with the following properties:
\begin{description}
  \item[I)] $f(\tilde{x},\tilde{u})=\tilde{x}$;
  \item[II)]$\tilde{x}\in\cX(\overline{x},\underline{N})$;
  \item[III)] $l(\tilde{x},\tilde{u})\leq \max\left(l(x^s,u^s),l(\overline{x},\overline{u})-\overline{\epsilon}\right)$.
\end{description}
\end{assumption}
The practical meaning of Assumption \ref{A:steady_state_sequence} is the following: for any fixed point $(\overline{x},\overline{u})$ in the set $\cF\times\U$, there exists another fixed point $(\tilde{x},\tilde{u})$ for the dynamics \eqref{E:system} (property I)), belonging to the set of reachable fixed points $\cX(\overline{x},\underline{N})$ (property II)). The value of the stage cost $l(\tilde{x},\tilde{u})$ is either equal to the minimal one among all fixed points, $l(x^s,u^s)$, or it is strictly lower, at least by $\overline{\epsilon}$, than $l(\overline{x},\overline{u})$ (property III)). We note that, if Assumption \ref{A:Feasible_set_1} holds, Assumption \ref{A:steady_state_sequence} is clearly satisfied at least for any fixed point $(\overline{x},\overline{u})\in\cF^s\times\U$, with $\underline{N}=N$ and $(\tilde{x},\tilde{u})=(x^s,u^s)$.\\
The modified MPC algorithm with generalized terminal state constraint is given below.\\
$\,$\\
\textbf{Algorithm 3} \emph{(Modified RH control with generalized terminal state constraint)
\begin{enumerate}
  \item (initialization) Select an arbitrarily small value of $\epsilon>0$ such that $2\epsilon\leq\overline{\epsilon}$, and select $\beta\geq\underline{\beta}(\epsilon)$.
      Given $x(0)\in\cF$, choose a value $\overline{l}(0)$ such that $(\overline{l}(0),x(0))\in\mathcal{S}$ and let $t=0$. Solve the FHOCP $\cP(\overline{l}(0),x(0))$; let $V^*(\overline{l}(0),x(0))$ be a solution. Apply to the system the control input $u(0)=v^*(0|0)$. Set $t=1$;
  \item set $\overline{l}(t)=\zeta(\overline{l}(t-1),x(t-1))$, solve the FHOCP $\cP(\overline{l}(t),x(t))$ by initializing the solver $\lambda$ with $\tilde{V}=\{v^*(1|t-1),\ldots,v^*(N|t-1),v^*(N|t-1)\}$; let $V^*(\overline{l}(t),x(t))$
  be a solution;
  \item if $l(x^*(N|t),v^*(N|t))>\overline{l}(t)-\epsilon$ and\\ $l(x^*(N|t),v^*(N|t))> l(x^s,u^s)+\epsilon$, then set $V^*(\overline{l}(t),x(t))=\tilde{V}$ and, consequently,\\ $(x^*(N|t),v^*(N|t))=(x^*_{N|t-1},v^*_{N|t-1})$;
  \item apply the control input $u(t)=\kappa(\overline{l}(t),x(t))$;
  \item  set $t=t+1$ and go to 2).
\end{enumerate}}

\begin{remark} Note that functions $\kappa(\overline{l}(t),x(t))$ and $\zeta(\overline{l}(t),x(t))$ used in Algorithm 3 are still the ones defined in \eqref{E:functions_implicit}, however the values of $v^*(0|t),\,x^*(N|t)$ and $v^*(N|t)$, needed for their evaluation, can be either the ones corresponding to the solution of problem $\cP(\overline{l}(t),x(t))$, or the ones corresponding to the control sequence $\tilde{V}$, depending on whether the condition $l(x^*(N|t),v^*(N|t))>\overline{l}(t)-\epsilon$ and $l(x^*(N|t),v^*(N|t))> l(x^s,u^s)+\epsilon$ is detected at step (3) of the algorithm. In virtue of Theorem \ref{T:perf_feas_set_s}-\textbf{a)}, such a condition may hold true only when $x(t)\notin\cF^s$, thus Algorithms 2 and 3 might behave differently only outside the feasibility set $\cF^s$.
\end{remark}
\begin{remark}\label{R:Theorem_conv_2} At step (3) of Algorithm 3, we use the tail of the previously computed optimal control sequence just for the sake of simplicity. One other option could be, at any time step $t$ such that the condition $l(x^*(N|t),v^*(N|t))>\overline{l}(t)-\epsilon$ and $l(x^*(N|t),v^*(N|t))> l(x^s,u^s)+\epsilon$ is detected,  to use another auxiliary control sequence, computed by solving a tracking optimization problem to reach the terminal state $(x^*(N|t-1),v^*(N|t-1))$ in minimum time.\end{remark}
$\,$\\
The next Theorem shows that the use of Algorithm 3 produces a sequence of terminal stage costs $\{l(x^*(N|t),v^*(N|t))\}_{t=0}^\infty$ that converges in finite time, within the arbitrarily small tolerance $\epsilon$, to the optimal value $l(x^s,u^s)$.\\
\begin{thm}\label{T:convergence_alg_mod}
Let Assumptions \ref{A:Feasible_set_1}-\ref{A:min_l} and \ref{A:steady_state_sequence} hold, and consider the closed-loop system obtained by applying Algorithm 3 with $N\geq\underline{N}$. Then, for any value of $x(0)\in\cF$ there exists a finite number of time steps $\overline{T}(x(0))$ such that:
\begin{equation}\label{E:convergence_thm}
l(x^*(N|\overline{T}(x(0))),v^*(N|\overline{T}(x(0))))\leq l(x^s,u^s)+\epsilon
\end{equation}
\end{thm}
\begin{pf}Consider any $x(0)\in\cF$. If $x(0)\in\cF^s$, then by Theorem \ref{T:perf_feas_set_s}-\textbf{a)} we have $l(x^*(N|0),v^*(N|0))\leq l(x^s,u^s)+\epsilon,\,\forall t\geq0$, hence \eqref{E:convergence_thm} also holds with $\overline{T}(x(0))=0$. If $x(0)\in\cF\setminus\cF^s$, then an optimal terminal state-input pair, $(x^*(N|0),v^*(N|0))$, is computed,  and the control input $u(0)=\kappa(\overline{l}(0),x(0))=v^*(0|0)$ is applied to the system. Consider now a generic time step $t>0$: assuming that $l(x^*(N|t),v^*(N|t))> l(x^s,u^s)+\epsilon$, the following two cases may occur.
\begin{itemize}
  \item[(a)] if $l(x^*(N|t),v^*(N|t))\leq \overline{l}(x(t))-\epsilon$, then, considering that $\overline{l}(x(t))=\zeta(\overline{l}(t-1),x(t-1))=l(x^*(N|t-1),v^*(N|t-1))$, the optimal terminal stage cost decreases at least by the quantity $\epsilon$:
      \[
      l(x^*(N|t),v^*(N|t))-l(x^*(N|t-1),v^*(N|t-1))\leq-\epsilon.
      \]
  \item[(b)] if $l(x^*(N|t),v^*(N|t))>\overline{l}(x(t))-\epsilon$, then at step (3) of Algorithm 3 the solution $V^*(\overline{l}(t),x(t))$ of $\cP(\overline{l}(t),x(t))$ is replaced by the tail of the previously computed optimal solution. Let us denote with $\tau$ the last time step at which the solution $V^*(\overline{l}(\tau),x(\tau))$ of $\cP(\overline{l}(\tau),x(\tau))$ was retained. Therefore, the control input at time $t$ is $u(t)=v^*(t-\tau|\tau)$, and the trajectory of the system evolves according to the optimal one predicted at time step $\tau$. The same procedure is carried out as long as case (b) holds true, until the state $x(\tau+N)=x^*(N|\tau)$ is eventually reached, which happens at most in $N$ time steps. 
      In virtue of Assumption \ref{A:steady_state_sequence}, since $N\geq\underline{N}$, once the fixed point $(x(\tau+N),u(\tau+N))=(x^*(N|\tau),v^*(N|\tau))$ has been reached, there exists at least one fixed point $(\tilde{x},\tilde{u})$, such that $\tilde{x}\in\cX(x(\tau+N),N)$ and $l(\tilde{x},\tilde{u})\leq\max\left(l(x^s,u^s),l(x^*(N|\tau),v^*(N|\tau))-\overline{\epsilon}\right)$. Hence, by Definition \ref{D:opt_achievable_cost} we have $\underline{l}(x(\tau+N))\leq l(\tilde{x},\tilde{u})\leq\max\left(l(x^s,u^s),l(x^*(N|\tau),v^*(N|\tau))-\overline{\epsilon}\right)$.
      Now, if $\max\left(l(x^s,u^s),l(x^*(N|\tau),v^*(N|\tau))-\overline{\epsilon}\right)=l(x^s,u^s)$, then $\underline{l}(x(\tau+N))=l(x^s,u^s)$ and, by Proposition \ref{P:opt_terminal}, we have $l(x^*(N|\tau+N),v^*(N|\tau+N))\leq l(x^s,u^s)+\epsilon$, thus the result \eqref{E:convergence_thm} holds true. If, on the contrary, $\max\left(l(x^s,u^s),l(x^*(N|\tau),v^*(N|\tau))-\overline{\epsilon}\right)=l(x^*(N|\tau),v^*(N|\tau))-\overline{\epsilon}$, then\\ $\underline{l}(x(\tau+N))\leq l(x^*(N|\tau),v^*(N|\tau))-\overline{\epsilon}$ and, again by Proposition \ref{P:opt_terminal}, we have\\ $l(x^*(N|\tau+N),v^*(N|\tau+N))\leq \underline{l}(x(\tau+N))+\epsilon\leq l(x^*(N|\tau),v^*(N|\tau))-\overline{\epsilon}+\epsilon$. Since $2\epsilon\leq\overline{\epsilon}$, the inequality  $l(x^*(N|\tau+N),v^*(N|\tau+N))\leq l(x^*(N|\tau),v^*(N|\tau))-\epsilon$ holds true, i.e. after at most $N$ time steps, the terminal stage cost decreases:\\$
      l(x^*(N|\tau+N),v^*(N|\tau+N))-l(x^*(N|\tau),v^*(N|\tau))\leq-\epsilon$
\end{itemize}
Summing up, while either cases (a) or (b) occur, i.e. as long as condition $l(x^*(N|t),v^*(N|t))> l(x^s,u^s)+\epsilon$ holds true,  the quantity $l(x^*(N|t+N),v^*(N|t+N))$ generally decreases (not strictly) with $t$. In particular, the decrease is guaranteed to be always at least equal to $-\epsilon$, and to take place at most every $N$ time steps:
\begin{equation}\label{E:worst_case_diff}
\begin{array}{ll}
&\forall j\geq1,\\
&l(x^*(N|jN),v^*(N|jN))\\
-&l(x^*(N|(j-1)N),v^*(N|(j-1)N))\leq-\epsilon.
\end{array}
\end{equation}
Moreover, due to Assumption \ref{A:fl_continuity} we have:
\begin{equation}\label{E:max_stage_diff}
\begin{array}{ll}
&l(x^*(N|0),v^*(N|0))-l(x^s,u^s)\\
\leq&\alpha_l(\|(x^*_{N|0},v^*_{N|0})-(x^s,u^s)\|).
\end{array}
\end{equation}
Equations \eqref{E:worst_case_diff}-\eqref{E:max_stage_diff} lead to the following result:
\[
\begin{array}{rl}
&l(x^*(N|jN),v^*(N|jN))-l(x^s,u^s)\\\leq&\alpha_l(\|(x^*(N|0),v^*(N|0))-(x^s,u^s)\|)-j\epsilon,
\end{array}
\]
hence when $j>\frac{\alpha_l(\|(x^*(N|0),v^*(N|0))-(x^s,u^s)\|)}{\epsilon}$ the condition $l(x^*(N|jN),v^*(N|jN))\leq l(x^s,u^s)+\epsilon$ is guaranteed to be satisfied. Therefore, we have
\[
l(x^*(N|\overline{T}(x(0))),v^*(N|\overline{T}(x(0))))\leq l(x^s,u^s)+\epsilon,
\]
where
\[
\overline{T}(x(0))=N\frac{\alpha_l(\|(x^*(N|0),v^*(N|0))-(x^s,u^s)\|)}{\epsilon}.
\]
Note that the pair $(x^*(N|0),v^*(N|0))$ is a function of the initial state $x(0)$ only, and thus also the quantity $\overline{T}(x(0))$ is.\hfill$\Box$
\end{pf}
\begin{remark}\label{R:Theorem_conv} According to Theorem \ref{T:convergence_alg_mod}, under the considered Assumptions, for any initial state $x(0)$ inside the feasibility set $\cF$, by using Algorithm 3 the stage cost of the terminal state-input pair converges to a value that is arbitrarily close to the optimal one, $l(x^s,u^s)$, after at most a finite number $\overline{T}(x(0))$ of time steps. Then, it can be noted that all the properties of Theorem \ref{T:perf_feas_set_s} hold true also if Algorithm 3 is used, for all time steps $t\geq\overline{T}(x(0))$.\end{remark}

\section{Examples}\label{S:Example}
\subsection{Linear-quadratic tracking MPC}\label{SS:example_1}
\begin{figure}[!hbt]
\centerline{
\begin{tabular}{c}
(a)\\
\includegraphics[bbllx=24mm,bblly=87mm,bburx=179mm,bbury=204mm,width=8.00cm,clip]{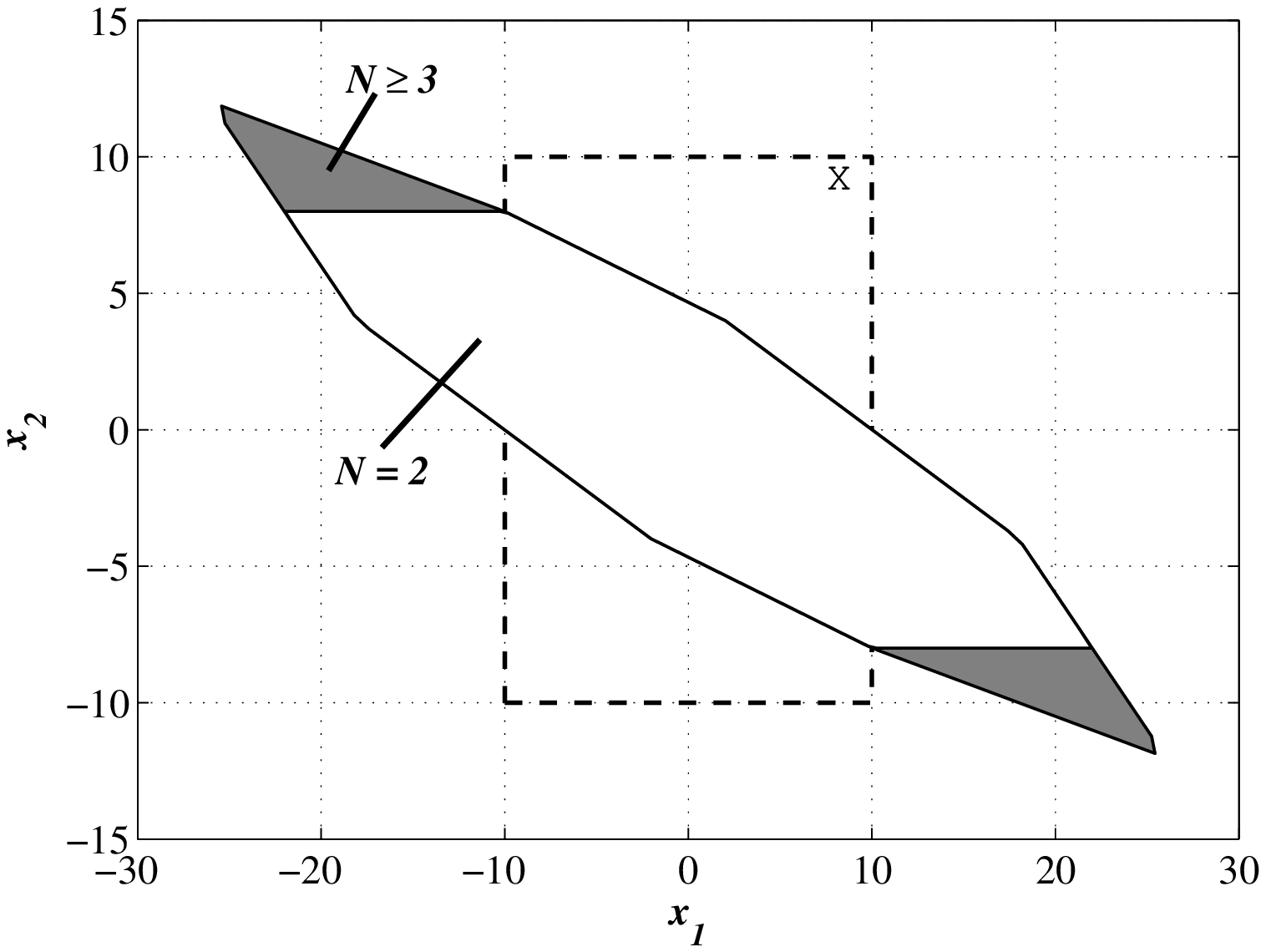}\\
(b)\\
\includegraphics[bbllx=24mm,bblly=87mm,bburx=179mm,bbury=204mm,width=8.00cm,clip]{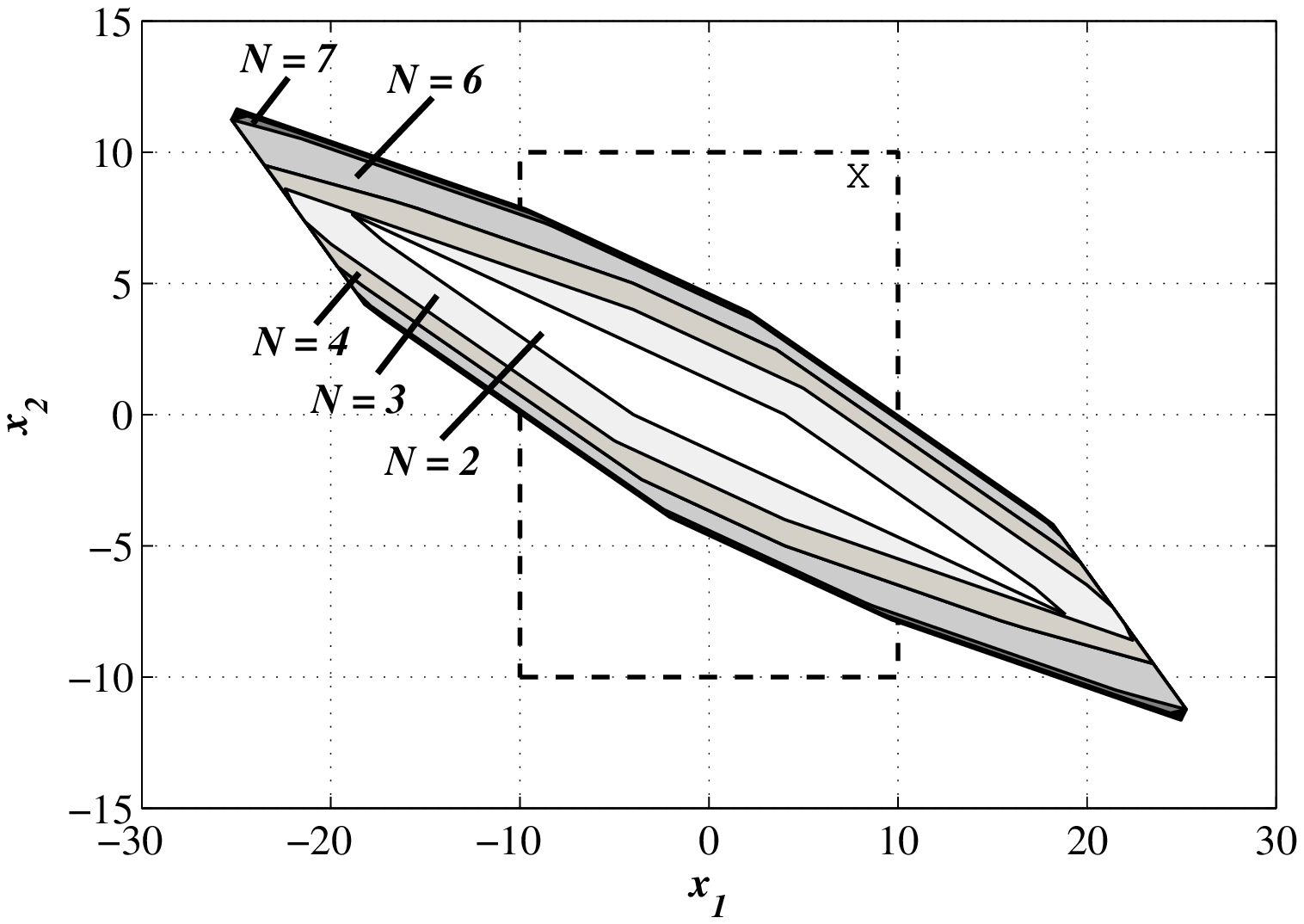}
\end{tabular}}
\caption{Linear quadratic tracking MPC. Feasibility sets (a) $\cF$  and (b) $\cF^s$ obtained with different prediction horizons $N$ and state constraint set $\X=\{x\in\R^2:\|x\|_\infty\leq10\}$.} \label{F:feas_example}
\end{figure}

\begin{figure}[!hbt]
\centerline{
\includegraphics[bbllx=24mm,bblly=87mm,bburx=179mm,bbury=204mm,width=8.00cm,clip]{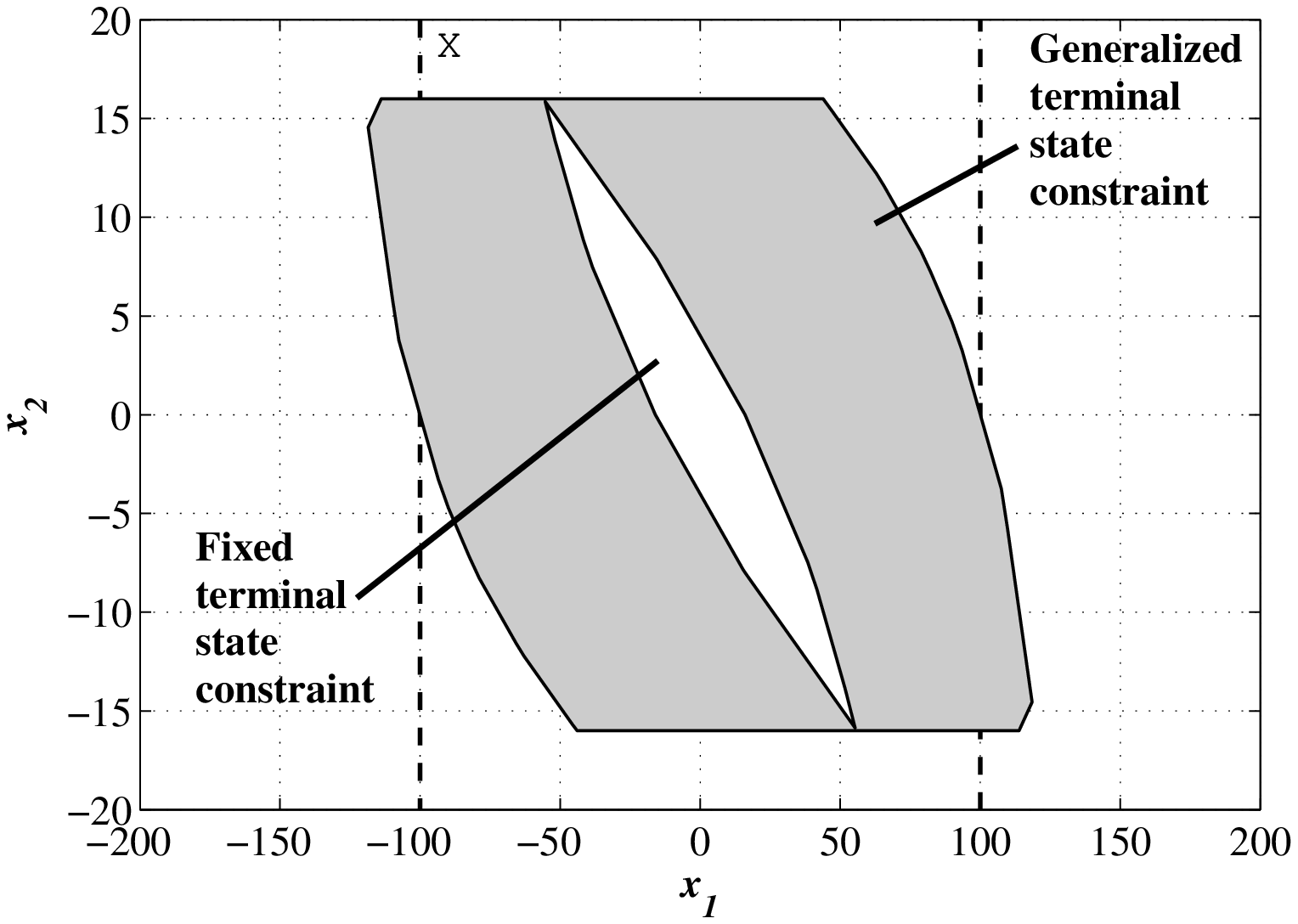}
}\caption{Linear quadratic tracking MPC. Feasibility sets $\cF$ (gray) and $\cF^s$ (white) obtained with prediction horizon $N=4$ and state constraint set $\X=\{x\in\R^2:\|x\|_\infty\leq100\}$.} \label{F:feas_example_enlargedX}
\end{figure}

\begin{figure}[!hbt]
\centerline{
\includegraphics[bbllx=24mm,bblly=90mm,bburx=175mm,bbury=202mm,width=8.00cm,clip]{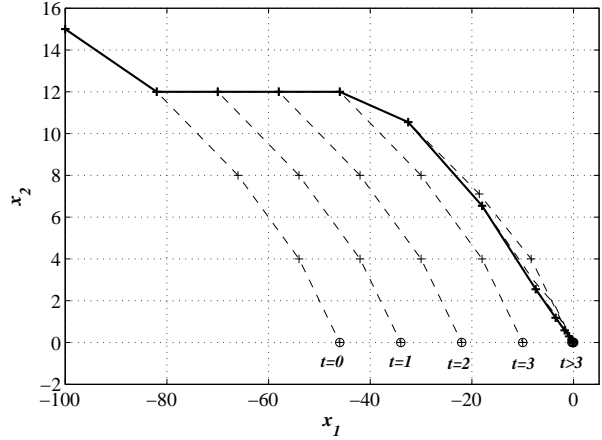}
}\caption{Linear quadratic tracking MPC. State trajectory obtained with the generalized terminal state constraint (solid), predicted trajectories (dashed), and predicted terminal states (circles). Initial condition $x(0)=[-100,15]^T$.} \label{F:x100_traj}
\end{figure}

\begin{figure}[!hbt]
\centerline{
\includegraphics[bbllx=20mm,bblly=85mm,bburx=177mm,bbury=205mm,width=8.00cm,clip]{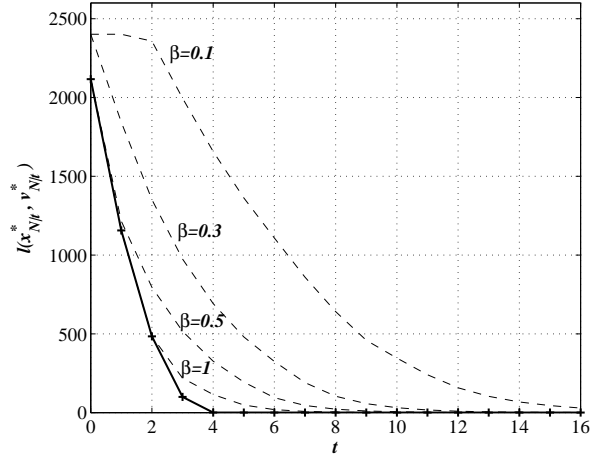}
}\caption{Linear quadratic tracking MPC. Course of the terminal stage cost $l(x^*(N|t),v^*(N|t))$ obtained with $\beta=1550$ (solid) and with $\beta=1,\,0.5,\,0.3$ and 0.1 (dashed).} \label{F:terminal_cost_traj}
\end{figure}

We consider the following linear system:
\[
x(t+1)=
\underbrace{\left[\begin{array}{rr}
1 & 1\\
0 & 1
\end{array}\right]}\limits_Ax(t)+\underbrace{\left[\begin{array}{rr}
1 & -1\\
-1 & 1
\end{array}\right]}\limits_Bu(t),
\]
with state constraint set $\X=\{x\in\R^2:\|x\|_\infty\leq10\}$ and input constraint set $\U=\{u\in\R^2:\|u\|_\infty\leq2\}$. The control problem is to regulate the state to the origin, hence a tracking MPC approach is used, with the following convex stage cost:
\[
l(x,u)=\|x\|_2+\|u\|_2,
\]
which satisfies Assumptions \ref{A:min_l} and \ref{A:tracking_stage_cost} with the Euclidean norm $\|(x,u)\|_2$ and $\underline{\alpha}_l(a)=a$. Assumption \ref{A:Feasible_set_1} is also satisfied for $N\geq2$, since the system is controllable and the state and input constraints contain the origin in their interiors. Assumption \ref{A:fl_continuity} is satisfied, considering the Euclidean norm, with $\alpha_f(a)=\gamma_f\,a$, $\gamma_f=\left\|\left[A\,B\right]\right\|_2=2.14$ and $\alpha_l(a)=\gamma_l\,a$, $\gamma_l=\|\nabla l\|_2=\sqrt{2}$, where $\nabla l$ is the gradient of $l$. Assumption \ref{A:solution_opt} holds, too, since problem $\cP$ is convex, due to the convexity of the stage cost and constraints, and the linearity of the model. The set of all admissible fixed points for this system is
$\{(x,u)\in\X\times\U:x_2=0,\,u_1=u_2\}$. Since the system is linear and controllable and the state and input constraints are convex,  it can be shown that, if $N\geq2$ (i.e. the state dimension), for any given fixed point $\overline{x}$ the set of reachable fixed points $\cX(\overline{x},N)$ \eqref{E:reachable_fixed_points} is convex. Then, due to the convexity of the stage cost and of the feasibility set, with a sufficiently small $\overline{\epsilon}$, for any given fixed point $(\overline{x},\overline{u})\in\cF\times\U$ there exists always a reachable fixed point $(\tilde{x},\tilde{u})\in\cX(\overline{x},N)\times\U$, such that $l(\tilde{x},\tilde{u})\leq\max(0,\,l(\overline{x},\overline{u})-\overline{\epsilon})$. Therefore, Assumption \ref{A:steady_state_sequence} is also valid.\\
We first compare the feasibility sets $\cF^s$ and $\cF$, obtained with different values of the prediction horizon $N$. Since the system dynamics and the constraints are linear, and the state constraints define a compact set, both $\cF^s$ and $\cF$ are polytopes. The result of the comparison, in Fig. \ref{F:feas_example},
clearly shows that the use of the generalized terminal state constraint yields an enlarged feasibility set, given the same horizon $N$. In particular, the feasibility set $\cF$ obtained with $N=2$ is already quite large, and the one obtained with $N=3$ corresponds to the maximal region of attraction for the origin, under the considered state and input constraints (see Fig \ref{F:feas_example}, (a)). A similar result can be obtained with the fixed terminal state constraint, but only with $N\geq7$ (see Fig \ref{F:feas_example}, (b)). More in details, the feasibility set $\cF^s$, obtained with a fixed terminal state constraint, corresponds to the $N$-step null controllable region of the system, subject to the considered state and input constraints, while, thanks to the generalized terminal state constraint, the feasibility set $\cF$ corresponds to the union of all of the $N$-step null controllable regions of the system, obtained by translating the origin to all of the feasible fixed points, again under the considered state and input constraints, hence it is larger than $\cF^s$. This aspect is even more evident with a larger state constraint set $\X$: in Fig. \ref{F:feas_example_enlargedX},
the case $\X=\{x\in\R^2:\|x\|_\infty\leq100\}$ is shown, with $N=4$ for both problems $\cP^s$ and $\cP$. It can be clearly noted that the set $\cF$ is equal, in this case, to the union of sets $\cF^s$ centered at all state values $\{x:x_2=0\}$, i.e. all fixed points for the system, limited only by the state constraint set $\X$. Indeed, in the absence of state constraints, the feasibility set $\cF$ would be unbounded. This example clearly shows that the proposed generalized terminal state constraint is able to remove the main drawback of using a fixed terminal state constraint, i.e. the need to use long prediction horizons in order to have a sufficiently large feasibility set, and it may also outperform, as far as the feasibility set is concerned, the two-mode approaches, in which it is still required for the state to reach in finite time a terminal set, which is positively invariant under a terminal control policy.\\
Considering the latter constraint set, we apply Algorithm 3 with a prediction horizon $N=4$. We obtain the value $\eta=150.08$ from \eqref{E:eta}, and we select $\epsilon=0.1$, hence the value $\underline{\beta}(\epsilon)=1500.8$. According to  Proposition \ref{P:opt_terminal}, we choose  $\beta=1550$. Fig. \ref{F:x100_traj} shows the closed-loop trajectory obtained from the initial condition $x(0)=[-100,15]^T$ (solid line), together with the trajectories predicted at each time step (dashed) and the related terminal states (circles).
According to Theorem \ref{T:convergence_alg_mod}, the terminal stage cost converges to zero in finite time, and specifically in 4 steps, as shown in Fig. \ref{F:terminal_cost_traj} (solid line), and, accordingly, the terminal state $x^*(N|t)$ converges to the origin (see Fig. \ref{F:x100_traj}), as stated in Theorem \ref{T:perf_feas_set_s}-\textbf{b)}. We note that the same result is obtained already with $\beta=50$, thus showing the conservativeness of the lower bound $\underline{\beta}(\epsilon)$, as anticipated in Remark \ref{R:optimal_terminal}.
A slower convergence is achieved with smaller values of $\beta$ (see Fig. \ref{F:terminal_cost_traj}, dashed lines). Finally, we also note that point (3) of Algorithm 3 is never applied in this example, i.e. at each time step the newly computed value of $l(x^*(N|t),v^*(N|t))$ improves the previous one, $l(x^*_{N|t-1},v^*_{N|t-1})$, by more than the selected value $\epsilon$, making Algorithm 3 equivalent to Algorithm 2.

\subsection{Economic MPC for an isothermal CSTR}\label{SS:example_2}
\begin{figure}[!hbt]
\centerline{
\includegraphics[bbllx=27mm,bblly=88mm,bburx=177mm,bbury=204mm,width=8.00cm,clip]{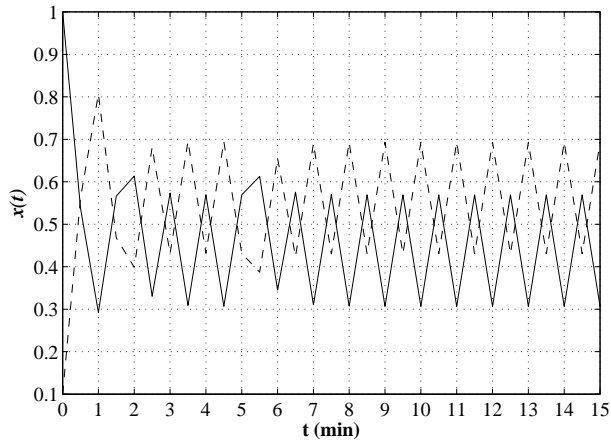}
}\caption{Isothermal CSTR. Course of the state variables $x_1(t)$ (solid) and $x_2(t)$ (dashed) obtained with the generalized terminal state constraint. Initial condition: $x(0)=[1,0.1]^T$.} \label{F:ese2_traj}
\end{figure}

\begin{figure}[!hbt]
\centerline{
\includegraphics[bbllx=21mm,bblly=88mm,bburx=177mm,bbury=204mm,width=8.00cm,clip]{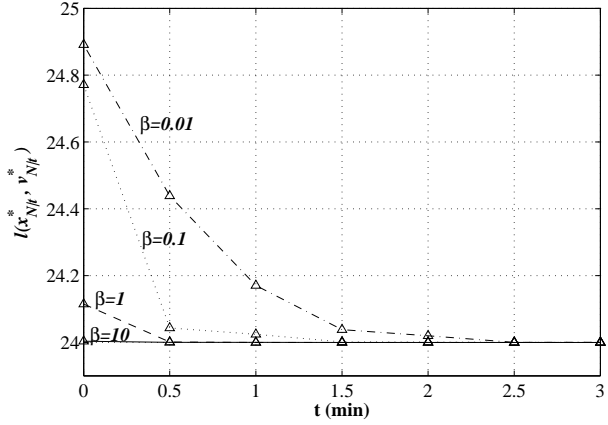}
}\caption{Isothermal CSTR. Course of the terminal stage cost $l(x^*(N|t),v^*(N|t))$ obtained with $\beta=10$ (solid), 1 (dashed), 0.1 (dot), and 0.01 (dash-dot). Initial condition: $x(0)=[1,0.1]^T$.} \label{F:ese2_traj_cost}
\end{figure}

We consider the model of an isothermal continuous stirring tank reactor (CSTR) from \cite{DiAR11}, with a single first-order reaction:
\[
C\rightarrow D.
\]
The state variables are the molar concentrations of $C$ and $D$, indicated as $x_1$ and $x_2$, respectively, and the input $u$ is the flow through the reactor, in l/min. The continuous-time dynamical equations of the model are:
\[
\begin{array}{l}
\dot{x}_1=\dfrac{u}{V_R}(c_{Cf}-x_1)-k_rx_1\\
\dot{x}_2=\dfrac{u}{V_R}(c_{Df}-x_1)+k_rx_1,
\end{array}
\]
where $V_R=10\,\,$l is the volume of the reactor, $c_{Cf}=1\,$mol/l and $c_{Df}=0\,$mol/l are the feed concentrations of $C$ and $D$, and $k_r=1.2\,\,$l/(mol min) is the rate constant. We restrict our attention to $x_1,x_2\in[0,1]\,$mol/l. The input constraint set is $\U=\{u\in\R:0\leq u\leq20\}$. The control objective is to minimize the following economic cost functional:
\[
l(x,u)=30-(2\,u\,x_2-\frac{1}{2}u),
\]
which attains its minimum $l(x,u)=0$ for $u=20$ and $\{x:x_2=1\}$.
The best steady state for the system, in terms of stage cost $l$, is $(x^s,u^s)=\left(\left[\begin{array}{c}0.5\\0.5\end{array}\right],12\right)$, and the corresponding cost is $l(x^s,u^s)=24$. The control sampling time is equal to $0.5\,$min. We use the generalized terminal state constraint with $N=12$. Fig. \ref{F:ese2_traj} shows the course of the states of the closed-loop system, starting from the initial condition $x(0)=[1,0.1]^T$.
The terminal stage cost, reported in Fig. \ref{F:ese2_traj_cost} with different values of $\beta$, rapidly converges to the optimal value, showing that the performance obtained with the generalized terminal state constraint are equivalent, in this case, to those achieved with an optimally-chosen fixed terminal state constraint.
As expected from Proposition \ref{P:opt_terminal}, the speed of convergence is higher with higher values of $\beta$ (see Fig. \ref{F:ese2_traj_cost}). After a first transient, a periodic state trajectory is obtained (see Fig. \ref{F:ese2_traj}),  cycling between the values $x=[0.57,0.43]^T$ mol/l and $x=[0.30,0.69]^T$ mol/l, and the corresponding input jumps between its extreme values, 0$\,$l/min and 20$\,$l/min, respectively. The corresponding asymptotic average cost $\overline{J}_\infty$ is equal to 21.14, i.e. better than the one associated to the optimal steady state $(x^s,u^s)$, as expected from the result \eqref{E:average_economic_cost_kappa_s} (see \cite{AnAR11}) and Theorem \ref{T:perf_feas_set_s}.

\subsection{Tracking MPC of an inverted pendulum system}\label{SS:example_3}
\begin{figure}[!hbt]
\centerline{
\includegraphics[bbllx=23mm,bblly=85mm,bburx=179mm,bbury=205mm,width=8cm,clip]{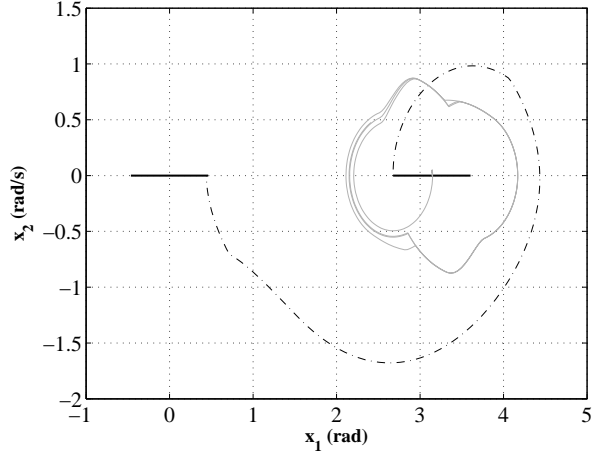}
} \caption{Inverted pendulum example: set of admissible steady states (solid lines), closed-loop trajectory obtained with horizon $N=60$ (gray lines) starting from $x(0)=[\pi,\,0]^T$ and state trajectory achievable in $N=141$ steps from the initial steady state $\overline{x}=[\pi-\arctan{(.5)},\,0]^T$ and input $\overline{u}=-0.5$.} \label{F:feas_pend}
\end{figure}

\noindent We consider the equations of motion of a pendulum written in
normalized variables (see \cite{AsFu00}):
\begin{equation}\label{E:pendolo}
\begin{array}{l} \dot{x}_1(t)=x_2(t)\\
\dot{x}_2(t)=\sin{(x_1(t))}-u(t)\cos{(x_1(t))}
\end{array}
\end{equation}
\noindent where the input constraint set $\mathbb{U}$ is:
\[
\mathbb{U}=\{u\in\mathbb{R}:|u|<0.5\}
\]
\noindent The state variables are the pendulum angular position
$x_1$ (modulo 2$\pi$) and angular speed $x_2$. We want to track the unstable set point
$[x_1,x_2,u]^T=[0,0,0]^T$, starting from the downright position $[x_1,x_2,u]^T=[\pi,0,0]^T$. The following discrete time model to be used in the nominal MPC design has been obtained by forward difference
approximation of (\ref{E:pendolo}):
\begin{equation}\label{E:pendolo_disc}
\begin{array}{l}
x_{1,t+1}=x_{1,t}+T_sx_{2,t}\\
x_{2,t+1}=x_{2,t}+T_s\,\left(\sin{(x_{1,t})}-u_t\cos{(x_{1,t})}\right)
\end{array}
\end{equation}
\noindent with sampling time $T_s=0.05\,$s. We choose a stage cost of the following form:
\[
l(x,u)=\left[\sin\left(x_1/2\right),\;x_2\right]Q\left[\begin{array}{c}\sin\left(x_1/2\right)\\x_2\end{array}\right]+u^2\,R,
\]
with $Q=\left[\begin{array}{cc}225&0\\0&1\end{array}\right]$ and $R=1$, and we choose $\beta=100$. The set of all admissible fixed points for the system is given by:
\[
\left\{(\overline{x},\overline{u}):\sin{(\overline{x}_1)}-\overline{u}\cos{(\overline{x}_1)}=0,\,\overline{x}_2=0,\,|\overline{u}|<0.5\right\},
\]
and it is depicted in Fig. \ref{F:feas_pend} (solid lines).
It can be noted that such a set is not connected, in particular its projection on the state space is given by:
\[
\left\{x:x_1\in[-\arctan(0.5),\,\arctan(0.5)]\,\cup\,[\pi-\arctan(0.5),\,\pi+\arctan(0.5)];\,x_2=0\right\}.
\]
In this situation, the use of a generalized terminal state constraint might not be able to drive the system state to the target $x^s=0$, starting from a steady state $\overline{x}$ such that $\overline{x}_1\in[\pi-\arctan(0.5),\pi+\arctan(0.5)]$, unless a sufficiently large horizon $N$ is chosen, as stated in our Assumption \ref{A:steady_state_sequence}. As an example, Fig. \ref{F:feas_pend} (gray line) shows the trajectory obtained with $N=60$, starting from the steady state/input $[x_1,x_2,u]^T=[\pi,0,0]^T$. It can be noted that the obtained trajectory approaches a limit cycle, confined in a set from which an admissible steady state $\overline{x}:\overline{x}_1\in[\pi-\arctan(0.5),\,\pi+\arctan(0.5)]$ is reachable within the considered horizon, yet an admissible steady state $\overline{x}:\overline{x}_1\in[-\arctan(0.5),\,\arctan(0.5)]$ is not reachable. In line with our result of Theorem \ref{T:perf_feas_set_s}, the average cost associated to such a limit cycle is equal to 195.89, while the best reachable steady state has an associated cost of 213.33.\\
A value of $N$ that satisfies Assumption \ref{A:steady_state_sequence} is $N=141$, as shown in Fig. \ref{F:feas_pend} (dash-dot line) by the related predicted trajectory starting from the initial steady state $\overline{x}=[\pi-\arctan{(.5)},\,0]^T$ and input $\overline{u}=-0.5$. However, note that since Assumption \ref{A:steady_state_sequence} is only sufficient for Theorem \ref{T:convergence_alg_mod} to hold, also lower values of $N$ might give the desired results. In particular, we show here the results with $N=100$. Fig. \ref{F:traj_1}(a) shows the obtained time courses of the state and input variables:
\begin{figure*}[tbh]
\centerline{\begin{tabular}{cc}
(a)&(b)\\
\includegraphics[bbllx=23mm,bblly=80mm,bburx=179mm,bbury=210mm,width=8cm,clip]{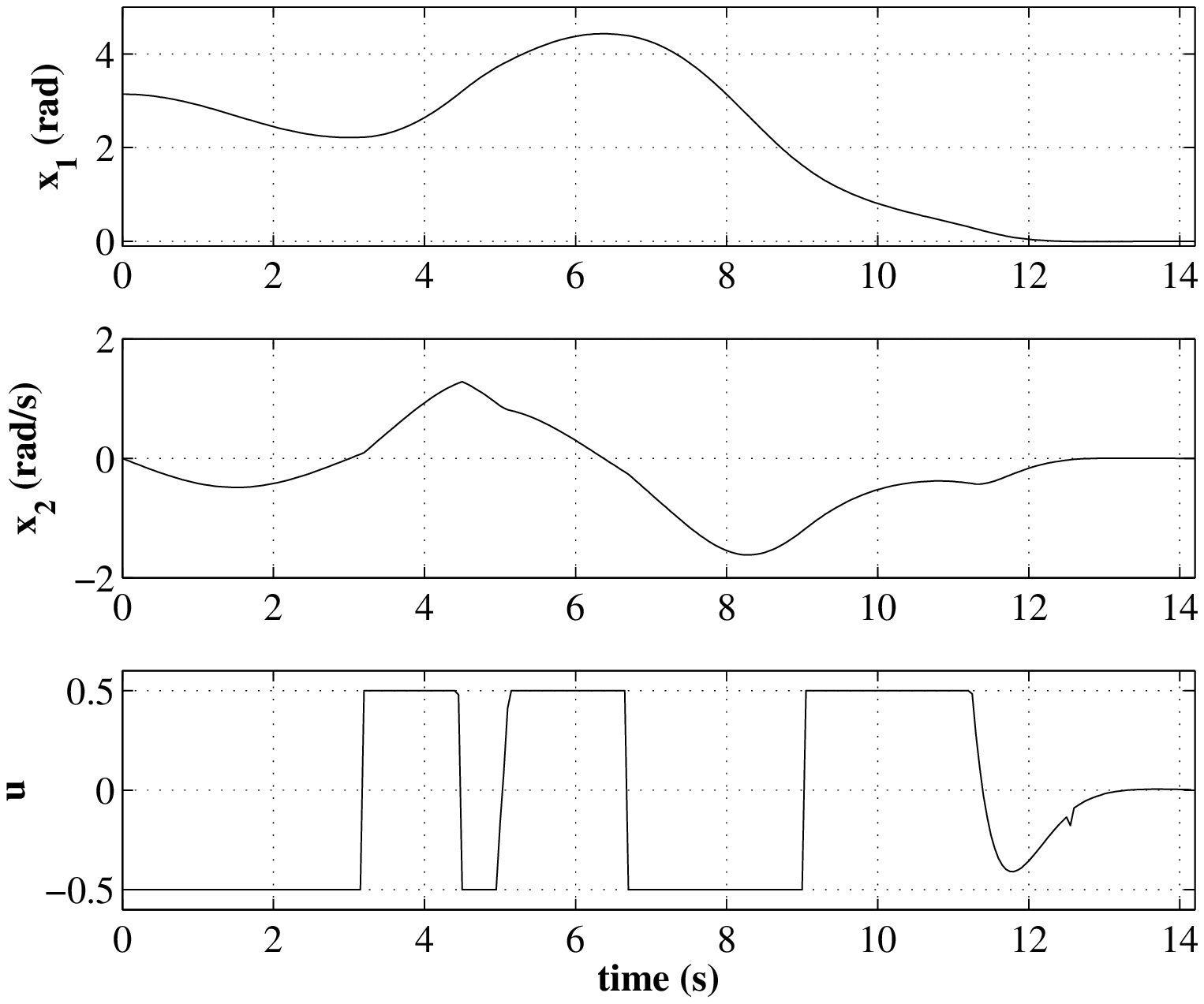}
&
\includegraphics[bbllx=18mm,bblly=80mm,bburx=180mm,bbury=209mm,width=8cm,clip]{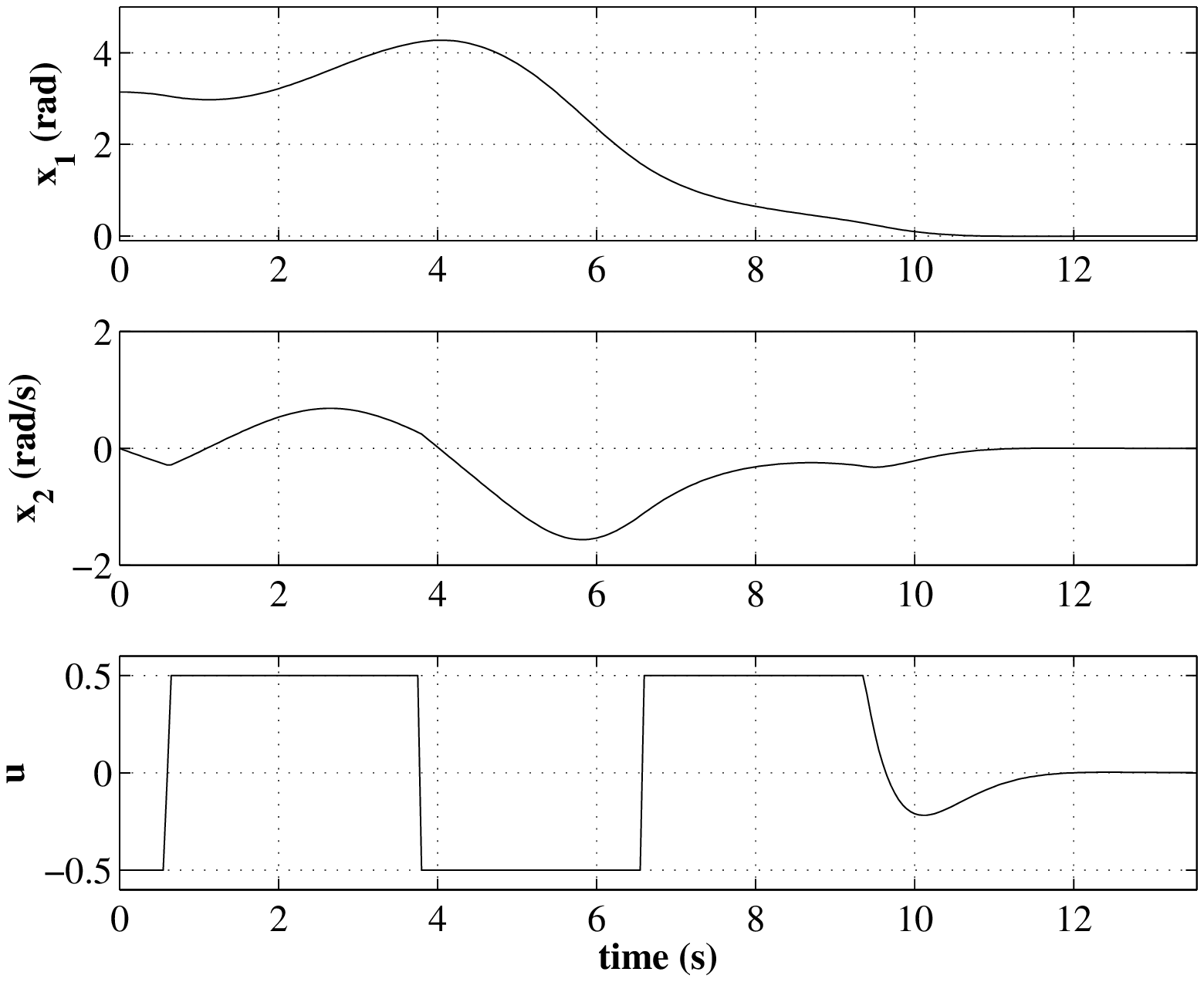}
\end{tabular}}
\caption{Inverted pendulum example: time courses of the states $x_1(t),\,x_2(t)$ and input $u(t)$ obtained with (a) the generalized terminal state constraint with $N=100$ and (b) a fixed, optimally chosen terminal state constraint with $N=200$.}
\label{F:traj_1}
\end{figure*}
\begin{figure*}[tbh]
\centerline{\begin{tabular}{cc}
(a)&(b)\\
\includegraphics[bbllx=18mm,bblly=77mm,bburx=179mm,bbury=212mm,width=8cm,clip]{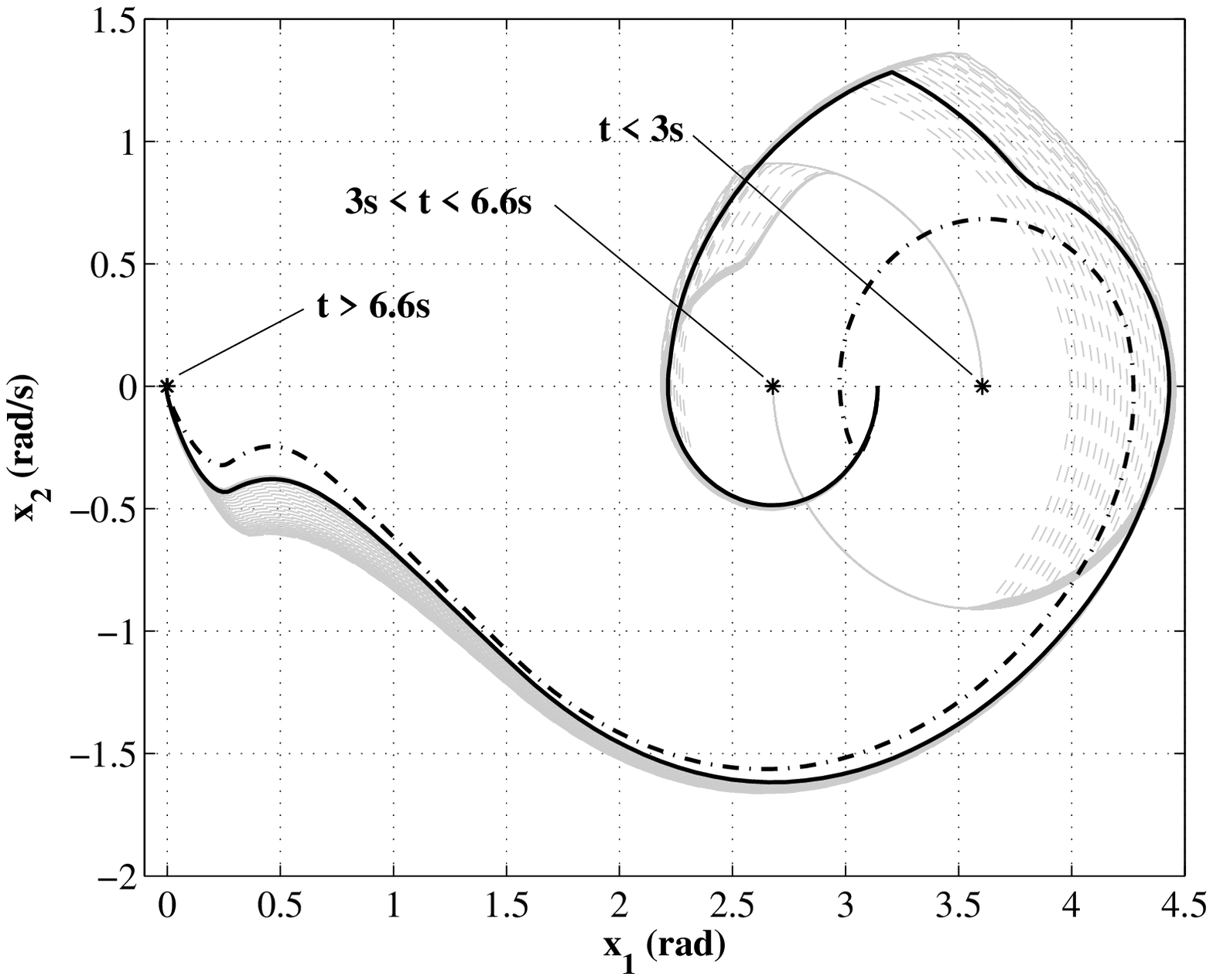}
&
\includegraphics[bbllx=18mm,bblly=83mm,bburx=180mm,bbury=209mm,width=8.5cm,clip]{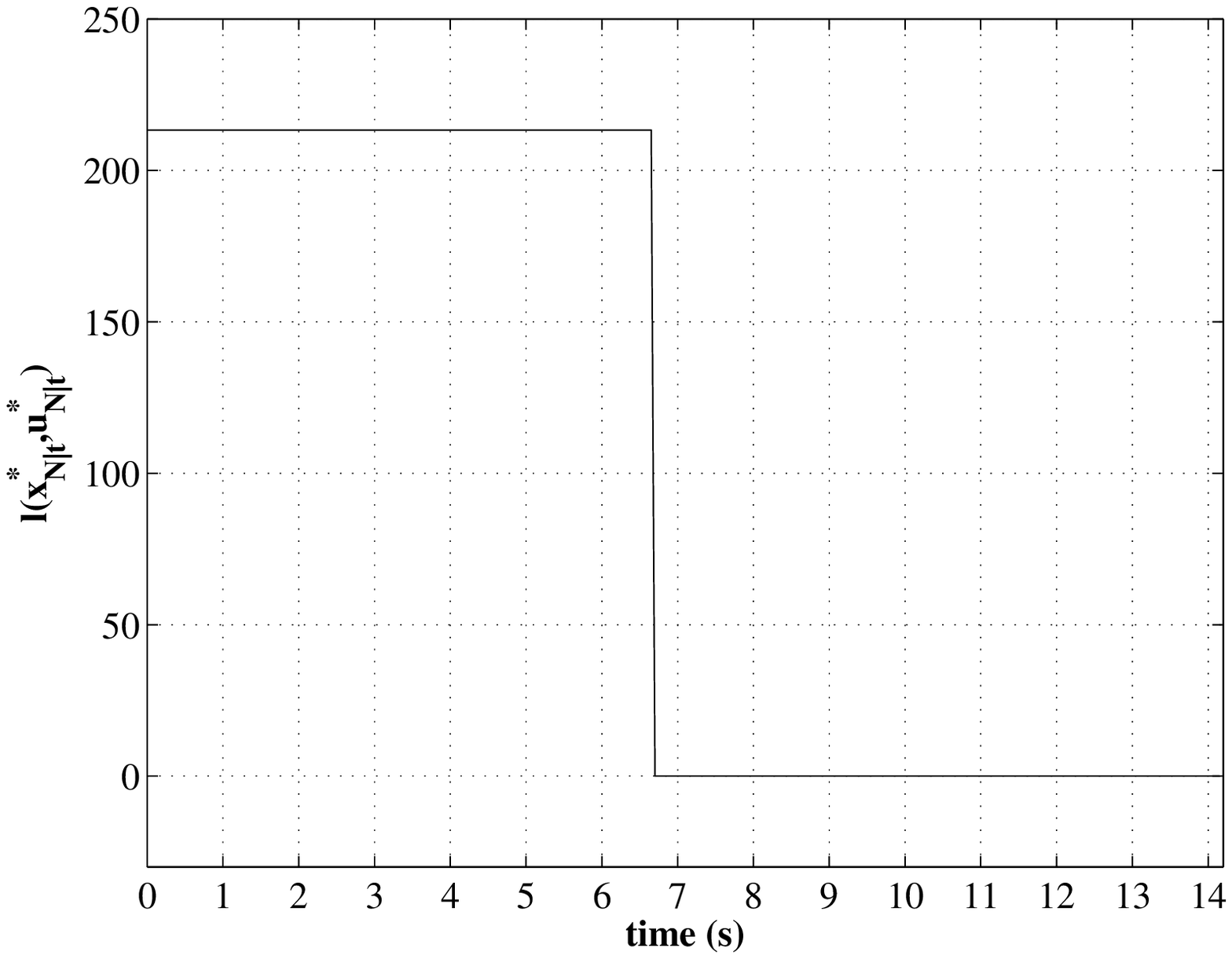}
\end{tabular}}
\caption{Inverted pendulum example: (a) closed loop state trajectory obtained with the generalized terminal state constraint (solid black line), with $N=100$, and trajectories predicted at each time step (gray dashed lines). Predicted terminal states are marked with `$*$'. Dash-dot black line: closed loop state trajectory obtained with a fixed terminal state constraint, and $N=200$. (b) Course of the stage cost associated to the terminal state input pair, $l(x^*(N|t),u^*(N|t))$.}
\label{F:traj_2}
\end{figure*}
the controller is able to swing up the pendulum in about 12.5$\,$s. The state trajectory in the phase plane is depicted in Fig. \ref{F:traj_2}(a) (solid black line), together with the trajectories predicted at each time step and the corresponding terminal steady states. In particular, it can be noted how the sequence of terminal state-input pairs, $(x^*(N|t),u^*(N|t))$, is approximately equal to $(3.60,0)$ for $t<3\,s$, then it jumps to $(2.67,0)$ for $t\in[3\,s,\,\,6.6\,s$ and finally converges to the target steady state, after about time 132 steps (i.e. 6.6$\,$s). These numerical values depend on the initial state, on the input constraints and on the chosen stage cost function and prediction horizon. Fig. \ref{F:traj_2}(b)   shows the same result in terms of terminal cost value, $l(x^*(N|t),u^*(N|t))$.

We also applied a tracking MPC law with a fixed terminal state constraint, i.e. $x(N|t)=0$. With the considered input constraint, the corresponding FHOCP \eqref{E:FHOCP} results to be unfeasible for horizons $N<200$, i.e. twice the one used with the generalized terminal state constraint. The trajectories obtained with the fixed terminal state constraint are shown in Figs. \ref{F:traj_2}(a) (dash-dot black line) and \ref{F:traj_1}(b): the controller is able to swing up the pendulum in about 11$\,$s. Therefore, this example confirms that 1. given the same prediction horizon, the use of a generalized terminal state constraint can give a feasibility set which is larger than that obtained with a fixed, optimally-chosen terminal state constraint, and 2. that the performance obtained with the generalized terminal state constraint, in terms of swing-up time, are very close to those achieved with a fixed terminal state constraint, but the required prediction horizon (i.e. computational effort) is much shorter.

\section{Conclusions}\label{S:Conclusions}
We investigated a generalized terminal state constraint for Model Predictive Control and proved that, under reasonable assumptions, the resulting closed-loop system has performance similar to those of MPC schemes with a fixed, optimally chosen terminal state constraint, while enjoying a larger feasibility set and lower computational complexity. These features have been highlighted through three examples, considering tracking MPC and economic MPC, and both linear and nonlinear systems.

\section*{Appendix}
\emph{Proof of equations \eqref{E:bound_cost_diff}-\eqref{E:eta}}. Consider equation \eqref{E:proof_prop_aux}.
The term $\beta[\underline{l}(x(t))-l(\hat{x},\hat{u})]$ is less than $-\beta\epsilon$ because $\beta>0$ and the pair $(\hat{x},\hat{u})$ is such that $l(\hat{x},\hat{u})>\underline{l}(x(t))+\epsilon$. The terms $[l(\overline{x}(i|t),\overline{v}(i|t))-l(\hat{x}(i|t),\hat{v}(i|t))],\,i\in[0,N-1]$ can be bounded on the basis of Assumption \ref{A:fl_continuity}, as follows:\\
$
\begin{array}{rl}
& \text{case: }i=0;\\
&l(\overline{x}(0|t),\overline{v}(0|t))-l(\hat{x}(0|t),\hat{v}(0|t))\leq |l(\overline{x}(0|t),\overline{v}(0|t))-l(\hat{x}(0|t),\hat{v}(0|t))|\leq
\alpha_l\|\overline{v}(0|t)-\hat{v}(0|t)\|)\\
=&\alpha_l\left(\sum\limits_{j=0}^{i}\alpha_f^{(i-j)}(\|\overline{v}(j|t)-\hat{v}(j|t)\|)\right);\\
\end{array}$\\
$
\begin{array}{rl}
& \text{case: }i=1;\\
&l(\overline{x}(1|t),\overline{v}(1|t))-l(\hat{x}(1|t),\hat{v}(1|t))\leq|l(\overline{x}(1|t),\overline{v}(1|t))-l(\hat{x}(1|t),\hat{v}(1|t))|\\
\leq&\alpha_l(\|(\overline{x}(1|t),\overline{v}(1|t))-(\hat{x}(1|t),\hat{v}(1|t))\|)\leq\alpha_l(\|\overline{x}(1|t)-\hat{x}(1|t)\|+\|\overline{v}(1|t)-\hat{v}(1|t)\|)\\
\leq&\alpha_l(\alpha_f(\|\overline{v}(0|t)-\hat{v}(0|t)\|))+\alpha_l(\|\overline{v}(1|t)-\hat{v}(1|t)\|)\\
\Rightarrow& l(\overline{x}(1|t),\overline{v}(1|t))-l(\hat{x}(1|t),\hat{v}(1|t))\leq
\alpha_l(\alpha_f(\|\overline{v}_{0|t}-\hat{v}_{0|t}\|))+\alpha_l(\|\overline{v}(1|t)-\hat{v}(1|t)\|)\\
=&\alpha_l\left(\sum\limits_{j=0}^{i}\alpha_f^{(i-j)}(\|\overline{v}(j|t)-\hat{v}(j|t)\|)\right);\\
&\ldots
\end{array}$\\
$
\begin{array}{rl}
& \text{generic }i\\
&l(\overline{x}(i|t),\overline{v}(i|t))-l(\hat{x}(i|t),\hat{v}(i|t))\leq\alpha_l\left(\sum\limits_{j=0}^{i}\alpha_f^{(i-j)}(\|\overline{v}(j|t)-\hat{v}(j|t)\|)\right).
\end{array}
$\\
Thus, the second term in \eqref{E:proof_prop_aux} can be bounded by $\eta=\alpha_l\left(\sum\limits_{j=0}^{i}\alpha_f^{(i-j)}\left(\max\limits_{\overline{v},\hat{v}\in\U}\|\overline{v}-\hat{v}\|\right)\right)$. 
\hfill$\Box$

\end{document}